\documentclass[11pt,a4paper]{article}
\usepackage{jheppub}
\usepackage[utf8]{inputenc}
\usepackage[english]{babel}

\usepackage{orcidlink}

\usepackage{amsfonts}

\usepackage{units}

\newcommand{\dx}{\mathrm{d}}
\newcommand{\CF}{C_{\mathrm{F}}}
\newcommand{\CA}{C_{\mathrm{A}}}
\newcommand{\TF}{T_{\mathrm{F}}}
\newcommand{\Nf}{n_{\mathrm{f}}}
\newcommand{\as}{\alpha_s}
\newcommand{\qbar}{\bar{q}}
\newcommand{\muf}{\mu_\text{F}}
\newcommand{\Mmax}{M_{\text{max}}}
\newcommand{\Nmax}{N_{\text{max}}}
\newcommand{\change}[1]{\textcolor{black}{#1}}

\title{A semi-analytical \boldmath{$x$}-space solution for parton evolution --- Application to non-singlet and singlet DGLAP equation}

\author{Juliane Haug\orcidlink{0009-0008-3076-1122},}
\author{Oliver Sch\"ule\orcidlink{0009-0007-8347-0081}}
\author{and Fabian Wunder\orcidlink{0009-0007-4136-7844}}

\affiliation{Institut f\"ur Theoretische Physik, Universit\"at T\"ubingen, \\
	Kepler Center for Astro and Particle Physics, \\ 
	Auf der Morgenstelle 14, D-72076 T\"ubingen, Germany}

\emailAdd{juliane-clara-celine.haug@uni-tuebingen.de}
\emailAdd{oliver.schuele@uni-tuebingen.de}
\emailAdd{fabian.wunder@uni-tuebingen.de}

\abstract{We present a novel semi-analytical method for parton evolution.
It is based on constructing a family of analytic functions spanning $x$-space which is closed under the considered evolution equation.
Using these functions as a basis, the original integro-differential evolution equation transforms into a system of coupled ordinary differential equations, which can be solved numerically by restriction to a suitably chosen finite subsystem.
The evolved distributions are obtained as analytic functions in $x$ with numerically obtained coefficients, providing insight into the analytic behavior of the evolved parton distributions.
As a proof-of-principle, we apply our method to the leading order non-singlet and singlet DGLAP equation.
Comparing our results to traditional Mellin-space methods, we find good agreement.
The method is implemented in the code \texttt{POMPOM} in \texttt{Mathematica} as well as in \texttt{Python}.}
\arxivnumber{2404.18667}

\begin{document}
\maketitle

\flushbottom

\section{Introduction}
The inner structure of hadrons, the bound states of quantum chromo dynamics (QCD), has been under investigation for the past 50 years \cite{Feynman1969,Bjorken1969,Drell1970,Cabibbo1978} and continues to be an active area of research to this day \cite{Gao2018,Ji2020,Ball2022}.
Since the low energy behavior of QCD is non-perturbative, the structure of hadrons cannot be calculated by perturbative methods.
Therefore, at present, parton distributions parametrizing the hadron structure have to be extracted from fits to experimental data \cite{DSSV2008,CTEQ2021,NNPDF2017}.
In parallel, there is an ongoing effort to calculate aspects of the hadron structure from first principles employing lattice methods \cite{TMD:2023}.
This approach is well suited for the prediction of various static properties of hadrons like the magnetic moment \cite{Martinelli:1982}, the axial charge \cite{Fucito:1982}, and form factors \citep{Martinelli:1987,Martinelli:1988a,Martinelli:1988b,Kronfeld:2019,Aoki:2022}.
Also (low) Mellin-moments of parton distributions are accessible on the lattice, which have the potential to constrain global parton distribution fits \cite{Lin:2018}.
For reviews on recent accomplishments regarding the extraction of parton distributions on the lattice see references \citep{Hagler2010,Constantinou2021a,Constantinous2021b}.

The distribution functions describing the hadron structure enter observables, e.g. structure functions of deep inelastic scattering (DIS), through \textit{factorization theorems} \cite{Collins1989,Stewart2010,Becher2015} who schematically have the form
	\begin{equation}
		\sigma_h(Q,x) \,=\, \sum_p \left(H_p\otimes 	f_{p/h}\right)(Q,x,\muf) \,+\, \mathcal{O}\!\left(\frac{\Lambda_\text{QCD}}{Q}\right) .
		\label{eq: factorization formula}
	\end{equation}
Here $\sigma_h(Q,x)$ is a hadronic observable depending on a hard scale $Q$, which is much larger than the non-perturbative QCD mass scale $\Lambda_\text{QCD}$, and other kinematic quantities denoted by $x$, e.g. transverse momenta or rapidities. 
$\sigma_h(Q,x)$ can be factorized into a partonic hard part $H_p$ and a parton-in-hadron distribution function $f_{p/h}$ up to corrections of higher orders in the scale $\Lambda_\text{QCD}/{Q}$.
The hard part only depends on short-distance physics which can be described perturbatively in terms of partons $p$.
The distribution functions $f_{p/h}$, which only depend on the universal, i.e. process-independent long-distance physics, can be viewed as probability densities of finding a parton $p$ in the hadron $h$.
They are determined by fitting to available data, see \cite{Forte2020} for a review on the state of the art method using neural networks.

Both quantities $H_p$ and $f_{p/h}$ depend on the factorization scale $\muf$ separating perturbative and non-perturbative regimes.
To all orders in perturbation theory in the -- also scale-dependent -- strong coupling $\as(\mu^2)$, the overall scale dependence would cancel in $\sigma_h$.
Truncating the series for $H_p$ at a fixed order however results in a residual dependence on $\muf$.
From demanding $0=\frac{\dx}{\dx\muf}\,\sigma_h$ and knowing the scale dependence of $H_p$, one can derive \textit{evolution equations} for the scale dependence of the distribution functions $f_{p/h}$, schematically
	\begin{equation}
		\frac{\dx f_{p/h}}{\dx \muf} \,=\, P \otimes f_{p/h} \,.
	\end{equation}
Note that even though the distribution functions are intrinsically non-perturbative objects, the evolution equations have perturbatively calculable kernels $P$.

The most basic non-perturbative distribution functions describing the hadronic structure are \textit{parton distribution functions (PDFs)} \cite{Owens1992}.
They parametrize the probability of finding a parton of longitudinal momentum fraction $\xi$ in a parent hadron.
Different PDFs exist for unpolarized \cite{Owens1992}, longitudinally polarized \cite{DSSV2008}, and transversely polarized hadrons \cite{Artru1990}.
Their evolution equation is the so-called \textit{Dokshitzer-Gribov-Lipatov-Altarelli-Parisi (DGLAP) equation} \cite{Dokshitzer1977,Dokshitzer1980,Gribov1972,Altarelli1977}, which will be discussed in more detail below.

More general parton distributions, whose evolution has been studied in the literature \cite{Rogers2016,Diehl2016}, are the so called \textit{transverse momentum dependent parton distributions (TMDs)} \cite{AngelesMartinez2015,TMD:2023} and \textit{generalized parton distributions (GPDs)} \cite{Diehl2003}.
Both carry information about the three dimensional structure of the hadron.
TMDs depend on the transverse momentum of the parton and are necessary to capture the physics in scattering regimes of low transverse momentum while GPDs carry information about the transverse position of partons within the hadron.

Other non-perturbative objects for which evolution equations play an important role include \textit{higher-twist correlation functions} \cite{Bukhvostov1983,Ji2015}.
They appear at higher orders of $\Lambda_\text{QCD}$ in the factorization eq.\,\eqref{eq: factorization formula}.
An example is the Efremov-Teryaef-Qiu-Sterman (ETQS) function $T_\text{F}$ \cite{Efremov1985,Qiu1991} whose evolution is of considerable interest in the literature \cite{Kang2009,Zhou2009,Vogelsang2009,Braun2009,Schaefer2012}.
Existing evolution codes for this case employ a discretization approach \cite{Pirnay:2013}.

In this paper, we only explicitly discuss the DGLAP evolution of PDFs.
The motivation behind this study is to investigate a semi-analytical method that may be useful for more involved equations like the ETQS evolution.
For PDF evolution, we do not strive to compete with existing well-established and very successful methods, but rather take the DGLAP equation as a proof-of-principle for our new method, where we can easily compare its advantages and drawbacks relative to existing methods along with its numerical performance.

There are quite a few solution methods for the DGLAP equation in the literature.
The common approaches can be broadly divided into two classes, Mellin-space methods and $x$-space methods.
Mellin methods employ the convolution structure of the DGLAP equation, which under Mellin transformation becomes a simple multiplication.
This allows for an analytic solution in Mellin-space.
The back transformation into $x$-space is usually performed numerically.
Publicly available algorithms based on this idea are \texttt{partonevolution} \cite{Weinzierl2002}, \texttt{QCD-PEGASUS} \cite{Vogt2005}, and \texttt{EKO} \cite{Candido2022}.
The $x$-space methods usually employ a discretization in $x$ together with a local or global interpolation ansatz to bring the DGLAP equation into the form of an ordinary differential equation.
Algorithms using this method include \texttt{QCDnum} \cite{Botje2011}, \texttt{APFEL} \cite{Bertone2014}, \texttt{HOPPET} \cite{Salam2009}, and \texttt{ChiliPDF} \cite{Diehl:2022}.
The latter has recently been generalized to encompass double parton distributions \cite{Diehl:2023}.
Instead of using a basis of interpolating functions one can also use a brute-force discretization in $\muf$ \cite{Miyama1996,Hirai1998}.
Yet another approach that has been explored in the literature is expanding the PDFs and evolution kernels in Laguerre polynomials in $\ln(x)$ \cite{Furmanski:1981}.
This transforms the integro-differential equation to a summation of a finite number of Laguerre coefficients.
The evolution of these coefficients can be determined recursively, a numeric implementation was done in \cite{Kobayashi:1995}.

Our method belongs to the class of $x$-space approaches.
However, instead of discretizing in $x$ we choose $\frac{\ln^m (x)}{m!}\, x^n$ as an overcomplete family of spanning functions for $x$-space which is constructed such that it is closed under DGLAP evolution.
This transforms the DGLAP equation into an ordinary differential equation for the coefficients.
Upon truncation to a finite subsystem, the differential equation can be solved by an ordinary matrix exponential, where the matrix in the exponential follows from a Magnus expansion \cite{Blanes_2009}.
The goal is to use spanning functions which capture the analytic behavior in as few functions as possible such that we can restrict the infinite function basis to a finite set for subsequent numerical treatment with only small errors.
We find that to achieve a relative accuracy at the $0.01\%$ level for DGLAP evolution the chosen family requires $\mathcal{O}(200)$ members.
The choice of spanning functions as well as the selection of a finite subset is anything but unique and might be refined in subsequent studies.

The novel idea behind our approach is to get a grasp on the analytic behavior of PDFs in $x$-space generated by DGLAP evolution.
This sets our method apart from previously applied $x$-space methods which use discretization in $x$ and interpolate the resulting PDFs.
We think that the approach presented in this work might help to get insight on distribution functions whose behavior is less well-known, such as the $T_F$ function.
There, the starting point would be to find a suitable basis which is closed under ETQS evolution.
Another potential use of our method is the calculation of derivatives of parton distributions, which can be relevant for the study of kinematic limits \cite{Lyubovitskij:2024civ}.
Due to the analytic form in $x$, differentiation is straightforward without accumulating uncertainties through numerical differentiation.
Furthermore, the proposed method decouples the evolution of the coefficient from the transformation between $x$-space and Mellin-space which only affects the basis.
Hence, it can be applied for the construction of Mellin-space expressions with a simple analytic structure.

This paper is organized as follows.
Section \ref{sec:DGLAP_equation_and_evolution_basis} summarizes basic notions of the DGLAP equation.
In section \ref{sec:Semi-analytical_x-space_method} we introduce our semi-analytical $x$-space method and apply it to the non-singlet and singlet DGLAP equations, before briefly discussing the relation to Mellin-space.
In section \ref{sec:Numerical_Analysis} we study the numerical performance of our method and compare the results to the Les Houches benchmark PDFs \cite{Giele:2002hx} as well as PDFs obtained using established Mellin-space methods.
Section \ref{sec:Conclusion} concludes our paper and gives a brief outlook on ETQS evolution.
Appendix \ref{app:Master_Integrals} collects the master integrals needed for applying our method to the leading order (LO) DGLAP equations, \change{appendix \ref{app: Differential equation for LO non-singlet coefficients} gives the differential equation for the non-singlet coefficients explicitly,} while appendix \ref{app:Input_distributions} lists the input distributions we use for numerical evolution.
This paper comes with \texttt{Mathematica} and \texttt{Python} implementations of the program \texttt{POMPOM} (Parton evolution with matrices: polynomials, logarithms, and more) which employ the presented method.
They can be found as ancillary files on the arXiv.
\change{Appendix \ref{app: Computation time} gives an estimate of the computation time of the code.}

\section{DGLAP equation and evolution basis}\label{sec:DGLAP_equation_and_evolution_basis}
In this section, we briefly introduce basic notions of the \change{DGLAP} equation \cite{Dokshitzer1977,Dokshitzer1980,Gribov1972,Altarelli1977}.
At LO, $f(x)$ is the probability density to find a parton $f$ carrying longitudinal momentum fraction $x$ of its parent hadron.
This simple interpretation does not strictly hold at higher orders, where the parton densities become renormalization- and factorization-scheme dependent.
Through factorization, the PDFs acquire a dependence on the scale $\mu$. 
How the PDFs change when the scale is varied, is described by the DGLAP equation.
An intuitive derivation of the DGLAP equation can be found in standard textbooks \cite{Ellis1996}.
More precise arguments based on the operator product expansion and the renormalization group \cite{Georgi1974,Gross1974} are given in \cite{Collins2011}.

For the physical basis of quark, antiquark and gluon distributions ($q_i$, $\qbar_i$, $g$), the DGLAP equation reads
	\begin{align}
		\frac{\dx}{\dx\ln\mu^2} \,
		\begin{pmatrix}
			q_i\!\left(\mu^2,x\right) \\
			g\!\left(\mu^2,x\right)
		\end{pmatrix}  =\, \frac{\as\!\left(\mu^2\right)}{2\pi} \sum_{q_j,\qbar_j} \int_x^1 \frac{\dx \xi}{\xi} \,
		\begin{pmatrix}
			P_{q_iq_j}(\xi) & P_{q_ig}(\xi) \\
			P_{gq_j}(\xi) & P_{gg}(\xi)
		\end{pmatrix}
		\begin{pmatrix}
			q_j\!\left(\mu^2,\frac{x}{\xi}\right) \\
			g\!\left(\mu^2,\frac{x}{\xi}\right)
		\end{pmatrix} .
	\end{align}
The evolution kernels (or \textit{splitting functions}) can be calculated as perturbative series in the QCD coupling constant $\as$,
	\begin{align}
		P_{ij} \,=\, P_{ij}^{(0)} \,+\, \frac{\as}{2\pi} P_{ij}^{(1)} \,+\, 	\left(\frac{\as}{2\pi}\right)^{2} P_{ij}^{(2)} \,+\, \ldots \,.
	\end{align}
Charge conjugation invariance yields
	\begin{align}
		P_{q_i q_j} \,=\, P_{\qbar_i \qbar_j}\,, \quad  P_{q_i \qbar_j} \,=\, P_{\qbar_i q_j} \,,
	\end{align}
while due to flavor symmetry, we have $P_{q_ig} \,=\, P_{qg}$, $P_{gq_j} = P_{qg}$, and also
	\begin{align}
		P_{q_i q_j} \,=\, \delta_{ij} P_{qq}^V \,+\, P_{qq}^S \,, \quad P_{q_i \qbar_j} \,=\, \delta_{ij} P_{q\qbar}^V \,+\, P_{q\qbar}^S \,.
	\end{align}
The flavor-diagonal valence splitting function $P_{qq}^V$ starts at order $\as^0$, while $P_{q\qbar}^V$ and the sea contributions $P_{qq}^S$, $P_{q\qbar}^S$ start at order $\as^1$. The difference $P_{qq}^S - P_{q\qbar}^S$ is of order $\as^2$ \cite{Moch2004}.

Instead of the physical basis, we write the DGLAP equation in terms of the non-singlet and singlet quark PDFs, which are defined as
	\begin{align}
		q_{\text{ns},\,ij}\!\left(\mu^2,x\right) \,&\equiv\, q_i\!\left(\mu^2,x\right) \,-\, q_j\!\left(\mu^2,x\right) ,
		\label{eq:Def_qNS}
		\\
		q_{\text{s}}\!\left(\mu^2,x\right) \,&\equiv\, \sum_{i=1}^{\Nf} \left[ q_i\!\left(\mu^2,x\right) + \qbar_i\!\left(\mu^2,x\right) \right] .
		\label{eq:Def_qS}
	\end{align}
Here, $\Nf$ is the number of active quark flavors participating in the QCD dynamics. 
Since the physical basis has $2 \Nf$ quark distributions, we need $2 \Nf -1$ linearly independent non-singlet distributions.
Note that a linear combination of non-singlet distributions is again a non-singlet distribution.
The advantage of this evolution basis is that the non-singlet distributions decouple from the gluon distribution under DGLAP evolution,
	\begin{align}
		\frac{\dx}{\dx\ln\mu^2}\, q_{\text{ns}}\!\left(\mu^2,x\right)  = \frac{\as\!\left(\mu^2\right)}{2\pi} \int_x^1 \frac{\dx \xi}{\xi}\, P_{qq}(\xi)\, q_{\text{ns}}\!\left(\mu^2,\frac{x}{\xi}\right) .
		\label{eq:NonSinglet_DGLAP}
	\end{align}
Since the quark splitting functions have some flavor dependence at higher orders, the evolution kernel of the non-singlet equation depends on the specific choice of non-singlet distribution.
The scale evolution of the singlet and gluon distributions is governed by
	\begin{align}
		\frac{\dx}{\dx\ln\mu^2}\,
		\begin{pmatrix}
			q_{\text{s}}\!\left(\mu^2,x\right) \\
			g\!\left(\mu^2,x\right)
		\end{pmatrix}  = \frac{\as\!\left(\mu^2\right)}{2\pi} \int_x^1 \frac{\dx \xi}{\xi}\,
		\begin{pmatrix}
			P_{qq}(\xi) & 2 \Nf P_{qg}(\xi) \\
			P_{gq}(\xi) & P_{gg}(\xi)
		\end{pmatrix}
		\begin{pmatrix}
			q_{\text{s}}\!\left(\mu^2,\frac{x}{\xi}\right) \\
			g\!\left(\mu^2,\frac{x}{\xi}\right)
		\end{pmatrix} ,
		\label{eq:Singlet_DGLAP}
	\end{align}
where here,
	\begin{align}
		P_{qq} \,=\, \left( P_{qq}^V \,+\, P_{q\qbar}^V \right) \,+\, \Nf \left(P_{qq}^S \,+\, P_{q\qbar}^S\right) .
	\end{align}
The LO evolution kernels are (\cite{Ellis1996}, chapter 4 and references therein)
	\begin{align}
		P^{(0)}_{qq}(\xi) \,&=\, \CF \left( \frac{1+\xi^2}{(1-\xi)_+} \,+\, \frac{3}{2} \delta(1-\xi)\right) ,
		\\
		P^{(0)}_{qg}(\xi) \,&=\, \TF \left( \xi^2 \,+\, (1-\xi)^2 \right) ,
		\\
		P^{(0)}_{gq}(\xi) \,&=\, \CF \left(\frac{1+(1-\xi)^2}{\xi}\right) ,
		\\
		P^{(0)}_{gg}(\xi) \,&=\, 2\,\CA  \left(\frac{\xi}{(1-\xi)_+} \,+\, \frac{1-\xi}{\xi} \,+\, \xi(1-\xi)\right) \,+\, \frac{11\,\CA \,-\, 4\,\TF\,\Nf}{6}\delta(1-\xi) \,.
	\end{align}
As usual, the plus-distribution appearing here is defined through its integral against an arbitrary test function $f(\xi)$ regular for $\xi \rightarrow 1$,
	\begin{equation}
		\int_0^1\dx\xi\, \frac{f(\xi)}{(1-\xi)_+} \,\equiv\, \int_0^1\dx\xi\, \frac{f(\xi)-f(1)}{1-\xi} \,.
	\end{equation}
The next-to-leading order (NLO) evolution kernels can also be found in \cite{Ellis1996}.
The complete DGLAP kernels are currently known to NNLO \cite{Vogt2004,Moch2004,Blumlein:2021enk}, approximate results for the fourth-order (N$^3$LO) quark-gluon splitting function $P_{gq}^{(3)}$ were published recently  \cite{falcioni2024fourloop}.
The running coupling $\alpha_s$ is known to N$^4$LO accuracy
\cite{Baikov:2017,Herzog:2017,Luthe:2017}, a review on the empirical and theoretical knowledge of $\alpha_s$ can be found in \cite{Deur2016}.

The momentum fractions carried by the individual partons must
add up to the full momentum carried by the hadron, which translates to the \textit{momentum sum rule}
	\begin{align}
		\int_0^1 \dx x\, x \left( q_{\text{s}}\!\left(\mu^2, x \right) \,+\,
		g\!\left(\mu^2, x \right) \right) \,=\, 1 \,.
		\label{eq:momentum_sum_rule}
	\end{align}
Quark number conservation implies the \textit{flavor sum rule} for valence distributions $q_{v,i} \equiv q_i - \qbar_i$
	\begin{align}
		\int_0^1\dx x\, q_{v,i}(x) \,=\, \text{number of valence quarks } i \,.
		\label{eq:quark_number_sum_rule}
	\end{align}
As these sum rules hold independent of $\mu$ \cite{Collins2011}, they can be used to check the quality of our evolution.

\section{Semi-analytical \boldmath{$x$}-space solution for parton evolution}\label{sec:Semi-analytical_x-space_method}

\subsection{General idea}
To solve a general integro-differential equation of the form
	\begin{equation}
		\frac{\dx}{\dx \mu} \mathbf{f}(\mu,\vec{x}) \,=\, \left( \mathbf{P}\otimes \mathbf{f} \right) (\mu,\vec{x}) \,,
		\label{eq:General_integro-differential_equation}
	\end{equation}
where $\mathbf{P} \otimes$ denotes an integral operator in $\vec{x}$-space, we propose the ansatz
	\begin{align}
		\mathbf{f}(\mu,\vec{x}) \,=\, \sum_m\, a_m(\mu)\, \mathbf{f}_m(\vec{x}) \,,
	\end{align}
where the $\mathbf{f}_m$ are a suitable set of spanning functions, while all $\mu$-dependence is contained in the coefficients $a_m$.
These coefficients are unknown and will be determined by transforming the original integro-differential equation into an ordinary differential equation for the coefficients,
	\begin{align}
		\frac{\dx}{\dx\mu} a_m(\mu) \,=\, \mathcal{P}_{mn}(\mu)\, a_n(\mu) \,.
		\label{eq:General_ODE_for_coefficients}
	\end{align}
The evolution matrix $\mathcal{P}$ is obtained by evaluating the integral in eq.\,\eqref{eq:General_integro-differential_equation} and collecting the result with respect to the spanning functions.
Once determined, the evolution matrix can be applied to any choice of initial condition.
Initial conditions in the chosen basis can be determined in several ways.
One possibility is a fit of the ansatz to the initial condition given in $x$-space. 
If an analytic form of the initial conditions is available, the coefficients in the chosen basis can also be determined by a direct expansion in this basis.
This is the method we will employ for the numerical study in section \ref{sec:Numerical_Analysis}, details of our expansion can be found in the accompanying code.
If the initial conditions are only known numerically or in case it is preferable for reasons of numerical accuracy or cost, one may employ a suitable interpolation akin to the method proposed in \cite{Diehl:2022}.

Eq.\,\eqref{eq:General_ODE_for_coefficients} is formally solved by a matrix exponential.
In case $\mathcal{P}(\mu)$ and $\mathcal{P}(\mu')$ do not commute, careful treatment is required.
Using the Magnus expansion discussed in the upcoming section \ref{sec:Solving the LO non-singlet and singlet DGLAP equation}, the solution can still be written in terms of an ordinary matrix exponential.
Restricting ourselves to a finite basis, the infinite set of differential equations can be approximated by a finite set, and the matrix exponential can be performed numerically.

Thus the main challenge is finding a suitable set of spanning functions.
The initial function $\mathbf{f}(\mu_0,\vec{x})$ should be well approximated by a small finite set of spanning functions.
The coefficients $a_m(\mu_0)$ of this expansion must be small enough, as otherwise small truncation errors in the evolution matrix $\mathcal{P}$ may cause large errors.
Additionally, the set of spanning functions must be closed under the considered integro-differential equation.
While this requirement will in principle be satisfied by a power series in $x$, a pure power series will often be insufficient as it can lead to large truncation errors.
For example, if inserting a pure power series into the integro-differential equation generates powers of $\ln(x)$, these logarithms should be included among the spanning functions, since they are only poorly approximated by a truncated power series.
This might result in the set of spanning functions to be overcomplete, so the decomposition of $\mathbf{f}$ will not be unique, which however does not present an issue.

To illustrate our method and demonstrate its numerical feasibility, we will apply it to the leading order non-singlet and singlet DGLAP equation in this work.
In principle, the proposed method is applicable for any evolution equation of the structure of eq.\,\eqref{eq:General_integro-differential_equation}.

\subsection{Solving the non-singlet and singlet DGLAP equation}\label{sec:Solving the LO non-singlet and singlet DGLAP equation}
Let us first consider the non-singlet DGLAP equation \eqref{eq:NonSinglet_DGLAP}. 
To solve it in $x$-space, we make the following ansatz,
	\begin{align}
		q_{\text{ns}}\!\left(\mu^2,x\right) =\, \sum_{m,n=0}^{\infty} a_{mn}\!\left(\mu^2\right) \frac{\ln^m(x)\, x^n}{m!} \,.
		\label{eq:NonSinglet_DGLAP_ansatz_for_PDF}
	\end{align}
The explicit logarithms in our overcomplete family of spanning functions $\frac{\ln^m(x)\, x^n}{m!}$ are well suited for accurately capturing the small $x$-behavior, the large $x$-behavior is accounted for by polynomials.
At LO, plugging this ansatz into eq.\,\eqref{eq:NonSinglet_DGLAP} yields
	\begin{align}
		&\sum_{m,n=0}^{\infty} \frac{\dx a_{mn}\!\left(\mu^2\right)}{\dx\!\ln\mu^2} \frac{\ln^m(x)\, x^n}{m!} 
		\,=\, \frac{\as\!\left(\mu^2\right)}{2\pi}\, \CF \sum_{m,n=0}^{\infty} a_{mn}\!\left(\mu^2\right)
		\nonumber \\
		&\times \left[\frac{1}{x}\, I_1^{m,n+1} \,+\, x\, I_1^{m,n-1} \,+\, 2\, I_2^{m,n} +\, 2\, \ln(1-x)\, \frac{\ln^m(x)\, x^n}{m!} \,+\, \frac{3}{2} \frac{\ln^m(x)\, x^n}{m!} \right],
		\label{eq:NonSinglet_DGLAP_I1_I2}
	\end{align}
where the master integrals $I_1$ and $I_2$ are listed in appendix \ref{app:Master_Integrals}.
The higher-order evolution kernels can be treated in a similar manner.
For example, the NLO evolution kernels contain $\ln(1\pm x)$, $\mathrm{Li}_2(x)$, which lead to the same master integrals $I_1$ and $I_2$ when expanded as a series in $x$ before integration.

Inserting the master integrals and collecting with respect to our spanning functions, we compare coefficients on both sides to read off a differential equation for the coefficients $a_{MN}\!\left(\mu^2\right)$ in matrix form.
Since we will need to truncate the series expansion of $\ln(1-x)$ when restricting ourselves to a finite subset of spanning functions, the largest relative errors are to be expected near $x=1$.

The differential equation for the coefficients of the LO non-singlet DGLAP equation is explicitly given in appendix \ref{app: Differential equation for LO non-singlet coefficients}.
Structurally, it takes the form
	\begin{align}
		\frac{\dx a_{MN}\!\left(\mu^2\right)}{\dx\!\ln\mu^2} \,=\, \frac{\as\!\left(\mu^2\right)}{2\pi}\, \mathcal{P}^{ij}_{MN}\, a_{ij}\! \left(\mu^2\right) .
		\label{eq:NonSinglet_DGLAP_for_coefficients}
	\end{align}
Here, $MN$ is the row index while $ij$ is the column index of the matrix $\mathcal{P}$.
This matrix equation is solved by
	\begin{align}
		a_{MN}\left(\mu^2\right) =\, \exp\!\left[ \left(\int_{\ln\mu_0^2}^{\ln\mu^2} \dx\!\ln\tilde{\mu}^2\, \frac{\as\!\left(\tilde{\mu}^2\right)}{2\pi} \right)  \mathcal{P}^{ij}_{MN} \right]  a_{ij}\!\left( \mu_0^2\right) .
		\label{eq:NonSinglet_DGLAP_coefficients_solution}
	\end{align}
Note that when taking higher-order corrections to the splitting functions into account ${\mathcal{P} = \mathcal{P}^{(0)} + \frac{\as}{2\pi} \mathcal{P}^{(1)} + \left(\frac{\as}{2\pi}\right)^2 \mathcal{P}^{(2)} + \ldots}$, where the different $\mathcal{P}^{(n)}$ do not necessarily commute.
Using the Magnus expansion \cite{Blanes_2009}, eq.\,\eqref{eq:NonSinglet_DGLAP_for_coefficients} can still be solved by an ordinary matrix exponential, such that
	\begin{align} 
		a_{MN}\!\left(\mu^2\right) =\, \change{\prod_{k=1}^K}\exp\!\left[ \Omega\!\left(\mu^2_{k},\mu_{k-1}^2\right) \right] a_{MN}\!\left(\mu_0^2\right) ,\quad\change{\text{with }\mu^2_{K}=\mu^2,}
	\end{align}
where $\Omega \,=\, \Omega_1 \,+\, \Omega_2 \,+\, \ldots$ with
	\begin{align}
		\Omega_1\!\left(\mu^2,\mu_0^2\right) &=\, \int_{\ln\mu_0^2}^{\ln\mu^2} \dx\!\ln\mu_1^2 \left(\as\!\left(\mu_1^2\right) \mathcal{P}^{(0)} \,+\, \as^2\!\left(\mu_1^2\right) \mathcal{P}^{(1)} \,+\, \as^3\!\left(\mu_1^2\right) \mathcal{P}^{(2)} \,+\, \mathcal{O}\!\left(\as^4\right) \right) ,
		\\
 		\Omega_2\!\left(\mu^2,\mu_0^2\right) &=\, \frac{1}{2} \int_{\ln\mu_0^2}^{\ln\mu^2} \dx\!\ln\mu_1^2 \int_{\ln\mu_0^2}^{\ln\mu_1^2} \dx\!\ln\mu_2^2\; \as\!\left(\mu_1^2\right) \as^2\left(\mu_2^2\right) \left[ \mathcal{P}^{(0)},\, \mathcal{P}^{(1)}\right] \,+\, \mathcal{O}\!\left(\as^4\right) .
	\end{align}
We see that starting at NNLO, that is $\mathcal{O}(\as^3)$, commutators of different $\mathcal{P}^{(n)}$ contribute.
Expressions for $\Omega_{n\geq 3}$, which contain nested commutators of the $\mathcal{P}^{(n)}$, can be found in \cite{Blanes_2009} and references therein.
Contributions due to $\Omega_3$ are at least of $\mathcal{O}(\as^4)$.
\change{The intermediate points $\mu_k$ are required if $\mu$ is outside the domain of convergence of the Magnus expansion \cite{Blanes_2009}.
	They have to be chosen such that omitted terms in the expansion are sufficiently suppressed, in other words such that the counting of powers of $\alpha_s$ in the integrand is not spoiled by a large integration domain.
	The optimal number of slices $K$ will have to be determined on a case-by-case basis balancing between required accuracy and computation time.
	A detailed discussion of a Magnus expansion approach towards a differential equation of the form of eq.~\eqref{eq:General_ODE_for_coefficients} can be found in \cite{Blanes_2009}, where its numerical performance is compared to standard Runge-Kutta methods.
	Of course the latter could also be used for solving eq.~\eqref{eq:NonSinglet_DGLAP_for_coefficients}.
}
We note that using the Magnus expansion in the context of the DGLAP equation has also been proposed recently in \cite{Simonelli:2024} \change{to determine an analytical solution in Mellin-space beyond leading order}.
	
The evolution of the singlet quark distribution, which was defined in eq.\,\eqref{eq:Def_qS}, is coupled to that of the gluon distribution and governed by eq.\,\eqref{eq:Singlet_DGLAP}.
To solve the singlet DGLAP equation, we make an ansatz analogous to eq.\,\eqref{eq:NonSinglet_DGLAP_ansatz_for_PDF},
	\begin{align}
		\begin{pmatrix}
			q_{\text{s}}\!\left(\mu^2,x\right) \\
			g\!\left(\mu^2,x\right)
		\end{pmatrix} =\, \sum_{m=0,\,n=-1}^{\infty} 
		\begin{pmatrix}
			a^{q_{\text{s}}}_{mn}\!\left(\mu^2\right) \frac{\ln^m(x)\, x^n}{m!} \\
			a^{g}_{mn}\!\left(\mu^2\right) \frac{\ln^m(x)\, x^n}{m!}
		\end{pmatrix} .
		\label{eq:Singlet_DGLAP_ansatz_for_PDF}
	\end{align}
Note that we included $x^{-1}$ in our set of spanning functions to capture the known, more strongly divergent, small-$x$ behavior of the gluon and sea-quark PDFs.
Plugging this ansatz into eq.\,\eqref{eq:Singlet_DGLAP}, we write the appearing integrals in terms of the master integrals $I_i$ defined in appendix \ref{app:Master_Integrals}.
As in the non-singlet case, we insert these master integrals and collect with respect to our spanning functions, which yields a differential equation for the coefficients,
	\begin{align}
		\frac{\dx}{\dx\ln\mu^2}\,
		\begin{pmatrix}
			a^{q_{\text{s}}}_{MN}\!\left(\mu^2,x\right) \\
			a^g_{MN}\!\left(\mu^2,x\right)
		\end{pmatrix}
		=\,
		\frac{\as\!\left(\mu^2\right)}{2\pi} \mathcal{P}_{MN}^{ij}
		\begin{pmatrix}
			a^{q_{\text{s}}}_{ij}\!\left(\mu^2,x\right) \\
			a^g_{ij}\!\left(\mu^2,x\right)
		\end{pmatrix} .
		\label{eq:Singlet_DGLAP_for_coefficients}
	\end{align}
At LO, this equation is solved by
	\begin{align}
		\begin{pmatrix}
			a^{q_{\text{s}}}_{MN}\!\left(\mu^2,x\right) \\
			a^g_{MN}\!\left(\mu^2,x\right)
		\end{pmatrix}
		= 
		\exp\!\left[ \int_{\ln\mu_0^2}^{\ln\mu^2} \dx\ln\tilde{\mu}^2 \frac{\as(\tilde{\mu}^2)}{2\pi} \mathcal{P}^{ij}_{MN} \right]
		\begin{pmatrix}
			a^{q_{\text{s}}}_{ij}\!\left(\mu_0^2,x\right) \\
			a^g_{ij}\!\left(\mu_0^2,x\right)
			\label{eq:Singlet_DGLAP_coefficients_solution} 
		\end{pmatrix} .
	\end{align}

We can now take PDFs at an initial scale $\mu_0^2$, parametrize them in the form of eqs.\,  \eqref{eq:NonSinglet_DGLAP_ansatz_for_PDF} and \eqref{eq:Singlet_DGLAP_ansatz_for_PDF}, and evolve them to any final scale $\mu^2$. While in principle, the matrices $\mathcal{P}$ have infinite dimensions, in practice the spanning functions $\frac{\ln^m(x)\, x^n}{m!}$ of our ansatz yield only small corrections for large $m,n$ (assuming their coefficients do not grow too fast). Therefore, we can cut off the infinite series in our original ansatz at some $\Mmax$, $\Nmax(m)$, such that the PDFs are parametrized in the form
	\begin{align}
		 \sum_{m=0}^{\Mmax} \sum_{n=\nu}^{\Nmax(m)} a_{mn}\!\left(\mu^2\right) \frac{\ln^m(x)\, x^n}{m!} \,,
		\label{eq:NonSinglet_DGLAP_ansatz_for_PDF_Cutoff}
	\end{align}
where $\nu = 0, -1$ in the (non-)singlet case.
For such a finite dimensional parametrization, the matrix exponentials in eqs.\,\eqref{eq:NonSinglet_DGLAP_coefficients_solution} and \eqref{eq:Singlet_DGLAP_coefficients_solution} can be solved numerically.
Note that we have some freedom to choose which finite subset of basis functions we keep.
Generally, the numerical performance will increase by adding more functions, however for a fixed basis size choosing a good cut-off, keeping ``important" basis functions in favor of others, will greatly enhance the numerical quality of results.

\subsection{Relation to Mellin-space}
For brevity of notation, we only consider non-singlet PDFs in this section, the observations also apply to the singlet and gluon PDFs.
The Mellin transform $\mathcal{M}$ with Mellin-space variable $s$ of eq.\,\eqref{eq:NonSinglet_DGLAP_ansatz_for_PDF} is
	\begin{align}
		\mathcal{M}\!\left[q_{\text{ns}}\right]\!\left(\mu^2,s\right) \,=\, \int_0^1\dx x\, x^{s-1}\, q_{\text{ns}}\!\left(\mu^2,x\right) =\,
		 \sum_{m,n=0}^{\infty} a_{mn}\!\left(\mu^2\right) \frac{(-1)^m}{(n+s)^{m+1}} \,.
		 \label{eq:Mellin-trafo in basis}
	\end{align}
As the Mellin transform only acts on the spanning functions and not on the coefficients, the coefficients $a_{mn}$ in Mellin-space are identical to those in $x$-space.
Hence, the scale evolution decouples from the switch between $x$-space and Mellin-space.
The spanning functions in Mellin-space, $\frac{(-1)^m}{(n+s)^{m+1}}$, are meromorphic functions of $s$ on $\mathbb{C}$, with poles on the real axis for $s=-n$.

These properties can for example be exploited to obtain a simple analytic form of the Mellin transform of a PDF given numerically in $x$-space by using our $x$-space spanning functions to fit the PDF in the form of eq.\,\eqref{eq:NonSinglet_DGLAP_ansatz_for_PDF_Cutoff}, and afterwards replacing the $x$-space with the Mellin-space spanning functions.
As we will see in the next section, the numerical error in the evolution resulting from the truncation in eq.\,\eqref{eq:Mellin-trafo in basis} can be controlled.

\section{Numerical analysis}\label{sec:Numerical_Analysis}
The proposed $x$-space method to solve eqs.\,\change{\eqref{eq:NonSinglet_DGLAP} and} \eqref{eq:Singlet_DGLAP} is exact for an infinite number of basis functions.
The key question is whether we can achieve sufficient accuracy for a realistic finite number of basis functions.
The goal of this chapter is \change{to establish} the numerical validity of the presented method along with its implementation in \texttt{POMPOM} and to determine a finite set of basis functions giving optimal results.
The criteria we will use to assess the quality of the evolution are a comparison with existing methods, consistency under back-evolution, \change{the magnitude of sum rule violations} from truncation effects, and convergence when increasing the number of basis functions.

As discussed above, we need to select a finite set of basis functions to solve the infinite systems of coupled differential equations \eqref{eq:NonSinglet_DGLAP_for_coefficients} and \eqref{eq:Singlet_DGLAP_for_coefficients} numerically.
As we will see, choosing different cut-offs has a sizeable effect.
Hence, we start the numerical investigation of our method by determining a good cut-off $\Mmax$, $\Nmax(m)$ for the parametrization in eq.\,\eqref{eq:NonSinglet_DGLAP_ansatz_for_PDF_Cutoff}.
Here, we use the Les Houches PDFs as a benchmark for the quality of our method \cite{Giele:2002hx}.

Having established the validity of our $x$-space method, we use it to evolve a full set of sufficiently realistic PDFs to various scales between $\mu^2 = \unit[\change{0.26}]{GeV^2}$ and $\mu^2 \,=\, \unit[10^8]{GeV^2}$.
Subsequently, we perform back-evolution as a consistency check and compare the $x$-space evolution to PDFs obtained using Mellin-space methods.
Last, we \change{analyze} the numerical convergence properties of our $x$-space method with increasing number of basis functions.

\subsection{Choice of basis cut-off}

	\begin{figure}
		\centering
		\includegraphics[width=0.8\textwidth]{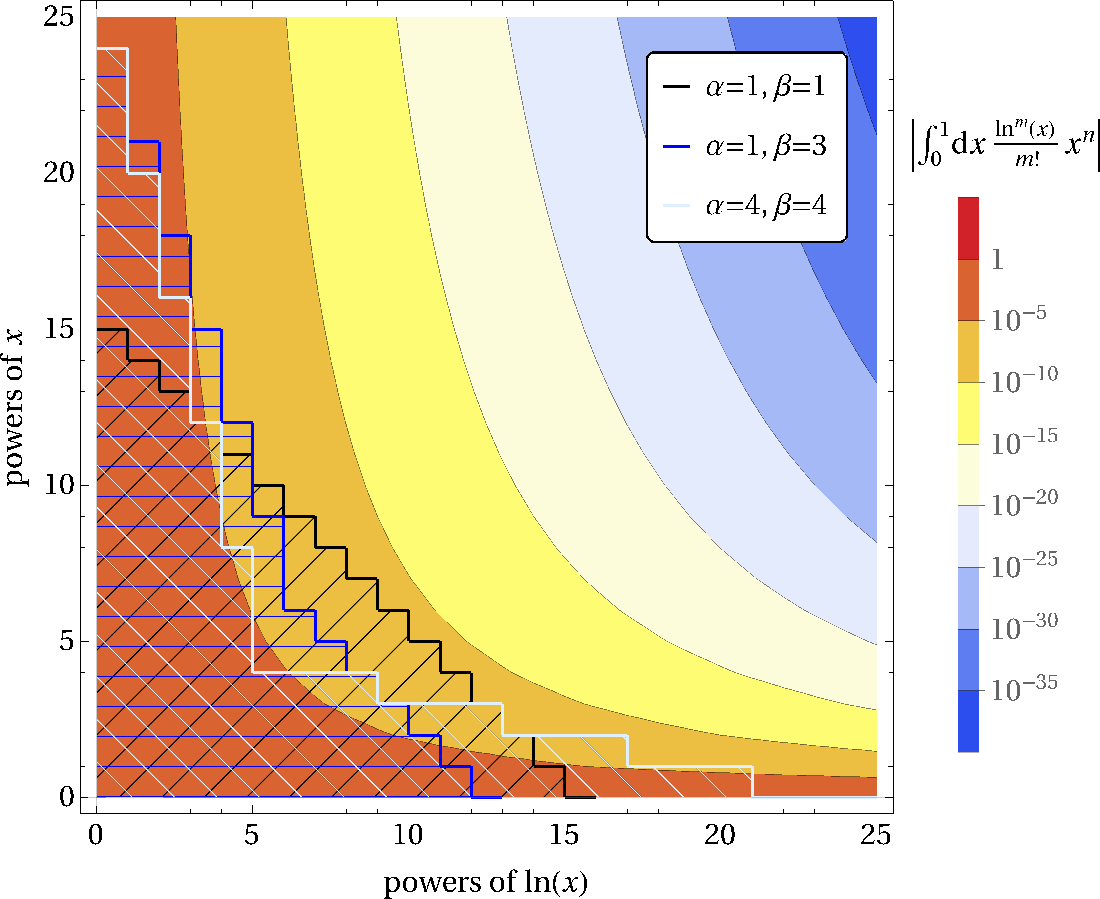}
		\caption{Plot of the contributions of different basis functions to the non-singlet normalization integral. The dashed areas indicate the included basis functions for different cut-offs of the form \eqref{eq: Linear cutoff function} with $150$ basis functions each.}
		\label{fig: Basis functions plot}
	\end{figure}
To establish the numerical validity of the proposed method we compare with the Les Houches benchmark PDFs \cite{Giele:2002hx}.
We observe that for a fixed total number of basis functions the quality of the agreement strongly depends on the choice of cut-off.
For given $\Mmax$, we parametrize $\Nmax(m)$ as 
	\begin{align}
		\Nmax^{\alpha,\beta}(m)=
		\left\lbrace\begin{array}{cl}
		\left\lfloor\frac{\Mmax-m}{\alpha}\right\rfloor & \quad\text{if }(\alpha+1)\,m>\Mmax ,
		\\
		\left\lfloor-\beta\,m+\frac{1+\beta}{1+\alpha}\Mmax\right\rfloor & \quad\text{else,}
		\end{array}
		\right.
		\label{eq: Linear cutoff function}
	\end{align}
where $\lfloor \ldots \rfloor$ denotes the floor function.
The two parameters $\alpha$ and $\beta$ allow for the inclusion of proportionately more pure polynomial or pure logarithmic terms and for varying the degree to which we want to exclude products.
Here, $-1/\alpha$ is the slope of the cut-off for powers of $\ln(x)$ greater than powers of $x$, $-\beta$ is the slope of the cut-off for powers of $x$ greater than powers of $\ln(x)$.
Cut-off functions with different $\alpha$ and $\beta$ are depicted in figure \ref{fig: Basis functions plot}.
We observe that the functions $\frac{\ln^m(x)}{m!} x^n$ rapidly become very small in the entire domain $0<x<1$ if both $m$ and $n$ increase simultaneously.
This behavior can also be seen in figure \ref{fig: Basis functions plot}, where we plotted $\left|\int_0^1\dx x \frac{\ln^m(x)}{m!} x^n\right|$ as a measure for the overall magnitude of the different basis functions.

\begin{figure}
		\centering
		\includegraphics[width=0.8\textwidth]{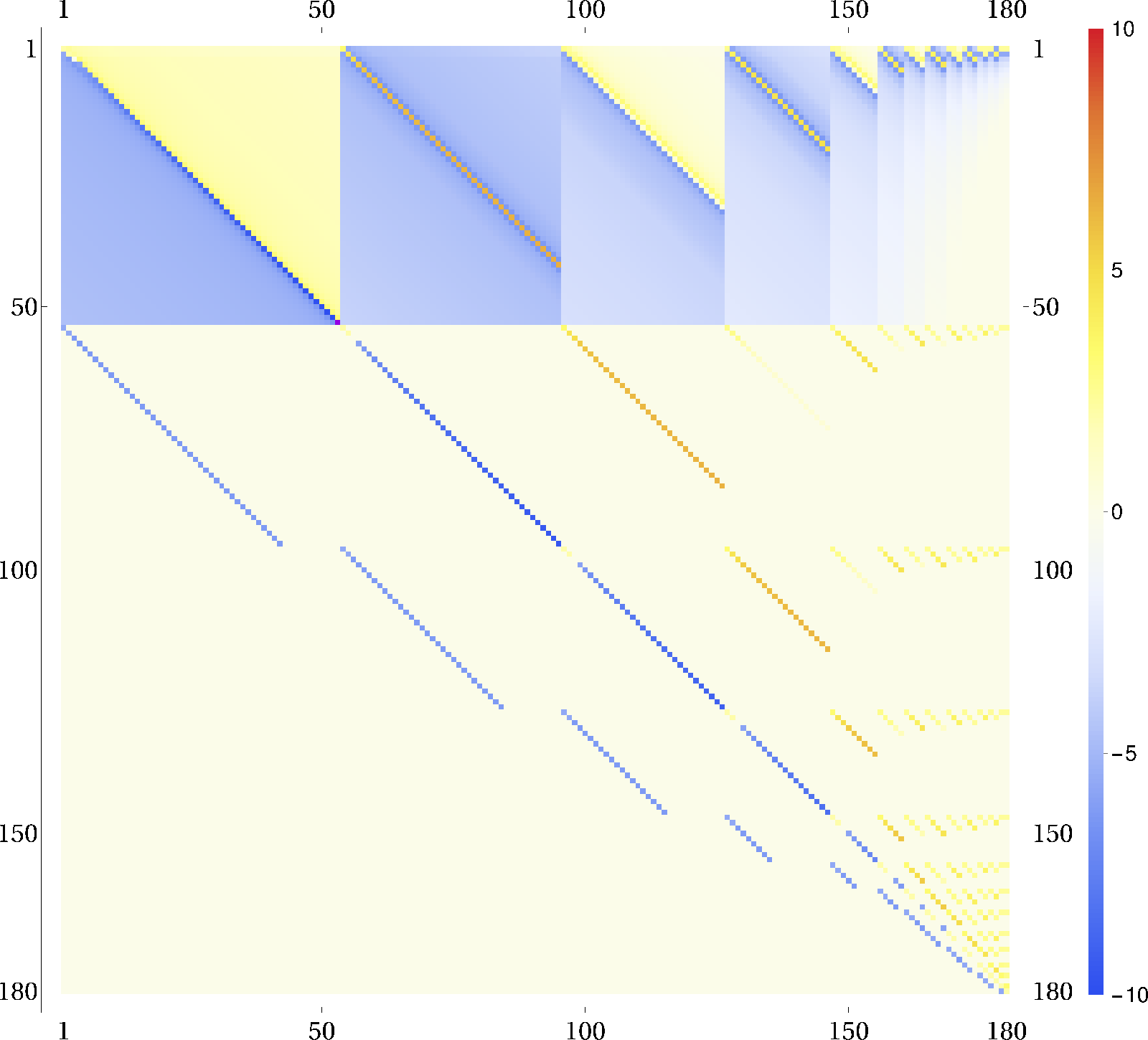}
		\caption{Magnitude of entries in the non-singlet evolution matrix for ${(\alpha,\beta) = (2,11)}$.
		The $180$ basis functions $\frac{\ln^m(x)}{m!} x^n$ are ordered lexicographically by $(n,m)$, i.e. $(1,x,x^2,\dots,\ln(x),x\,\ln(x),\cdots)$.
		The matrix is sparse except for the coupling to logarithm-free terms in the uppermost rows.
		Most of the entries have magnitude smaller than $1$.
		Except for some densely populated ``diagonals", the matrix elements are getting small when approaching the cut-offs.}
		\label{fig: Matrix plot}
	\end{figure}

However, the optimal cut-off has to take into account not only the magnitude of the basis functions but also the magnitude of the coefficients.
These are determined from the initial conditions and the respective elements of the evolution matrix.
A typical evolution matrix for \change{a particular} choice of total number of basis functions and cut-off is depicted in figure \ref{fig: Matrix plot}.

We observe that the matrix is most densely populated for pure polynomial terms coupling to pure polynomials.
With increasing powers of logarithms the coefficients coupling to pure polynomials decrease.
The rest of the matrix, which couples terms with at least one logarithm to other terms with logarithms is populated quite sparsely.
Only some ``diagonals" retain non-zero values.
The vast majority of matrix elements has values between $-1$ and $1$ depicted light blue respectively light yellow in figure \ref{fig: Matrix plot}.
Only on a few diagonals, due to the presence of harmonic numbers in the master integrals, terms grow logarithmically to values of the order of $\pm 10$.
Away from these diagonals the values decrease towards the cut-off.
These properties can also be read off from the differential equation for the non-singlet coefficients $a_{MN}$ in analytic form given in appendix \ref{app: Differential equation for LO non-singlet coefficients}.

The key observation which allows for restricting the infinite system to a finite matrix is that the coefficients do not grow strongly.
Overall, the numerical performance of the finite system may be spoiled by three sources:
	\begin{itemize}
		\item omitting ``large" basis functions,
		\item omitting matrix elements which couple strongly to the initial condition,
		\item large coefficients in the initial conditions or a very large number of non-negligible coefficients in the initial conditions.
	\end{itemize} 
Note that rescaling the basis functions can shuffle around the overall factors between the three.
The only way to avoid large coefficients in the initial conditions is choosing suitable spanning functions, the other two determine what finite cut-off is sensible.

Looking at figures \ref{fig: Basis functions plot} and \ref{fig: Matrix plot}, we see that choosing a finite cut-off drops some basis functions which are not per se small and also omits some not inherently small matrix elements.
However, we will nevertheless obtain a precise evolution provided that the omitted basis functions do not couple strongly to the initial conditions.
This claim is what needs to be checked numerically in the following.

\subsection{Comparison to Les Houches benchmark tables}
\begin{figure}
		\centering
		\vspace{-2cm}
		\includegraphics[width=1\textwidth]{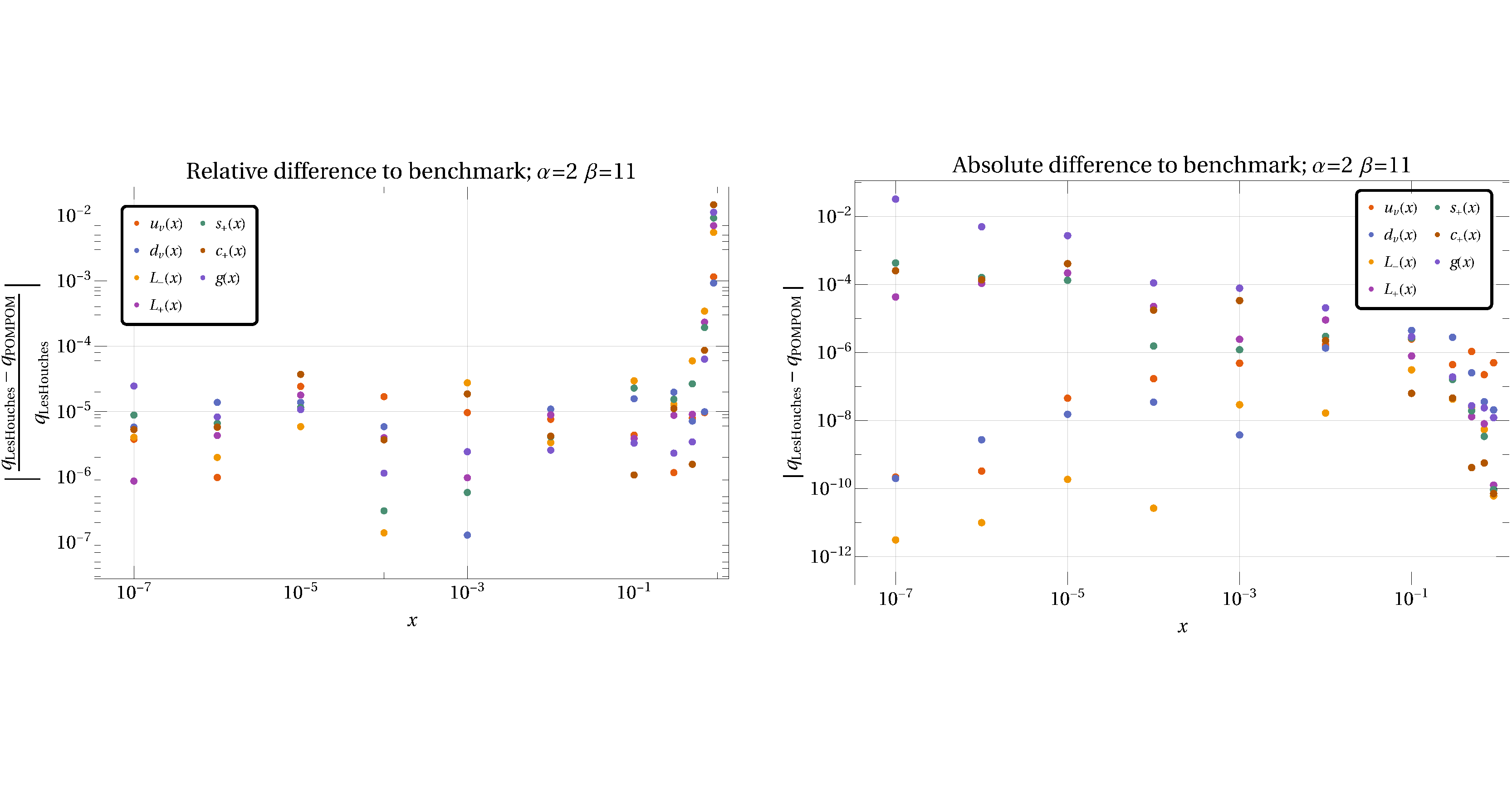}
		\vspace{-2cm}
		\caption{\change{Comparison between \texttt{POMPOM} with 200 basis functions and the Les Houches benchmark table for an evolution from $\mu_0^2=\unit[2]{GeV^2}$ to $\mu_\text{f}^2=\unit[10^4]{GeV^2}$ in a fixed flavor number scheme with $\Nf=4$.}}
		\label{fig: Comparison with Les Houches}
	\end{figure}
\change{We start the investigation of the numerical importance of the basis cut-off by comparing \texttt{POMPOM} against the \change{Les Houches} benchmark tables \cite{Giele:2002hx}.
These are commonly used for checking PDF evolution codes \cite{Candido2022}.
In figure \ref{fig: Comparison with Les Houches} we show the results of the comparison between \texttt{POMPOM} with 200 basis functions and the Les Houches benchmark table for an evolution from $\mu_0^2=\unit[2]{GeV^2}$ to $\mu_\text{f}^2=\unit[10^4]{GeV^2}$ in a fixed flavor number scheme with $\Nf=4$.
The \texttt{POMPOM} cut-off parameters are chosen as $(\alpha,\beta)=(2,11)$, this choice will be discussed in more detail below.
In the figure on the right we observe that for $x\leq 0.5$ the relative difference is less than $10^{-4}$, for all $x\leq 0.3$ there is agreement in all significant digits.
For the largest values of $x$ the relative deviations increase due to the smallness of the PDFs near  near $x=1$.
As the figure on the right demonstrates, the absolute error for $x\geq 0.5$ is below $10^{-6}$.
Hence we see that even though the cut-off on basis functions in the \texttt{POMPOM} method may lead to large relative deviations in the vicinity of $x=1$, the absolute error is under control.
}

		\begin{figure}
		\centering
		\includegraphics[width=\textwidth]{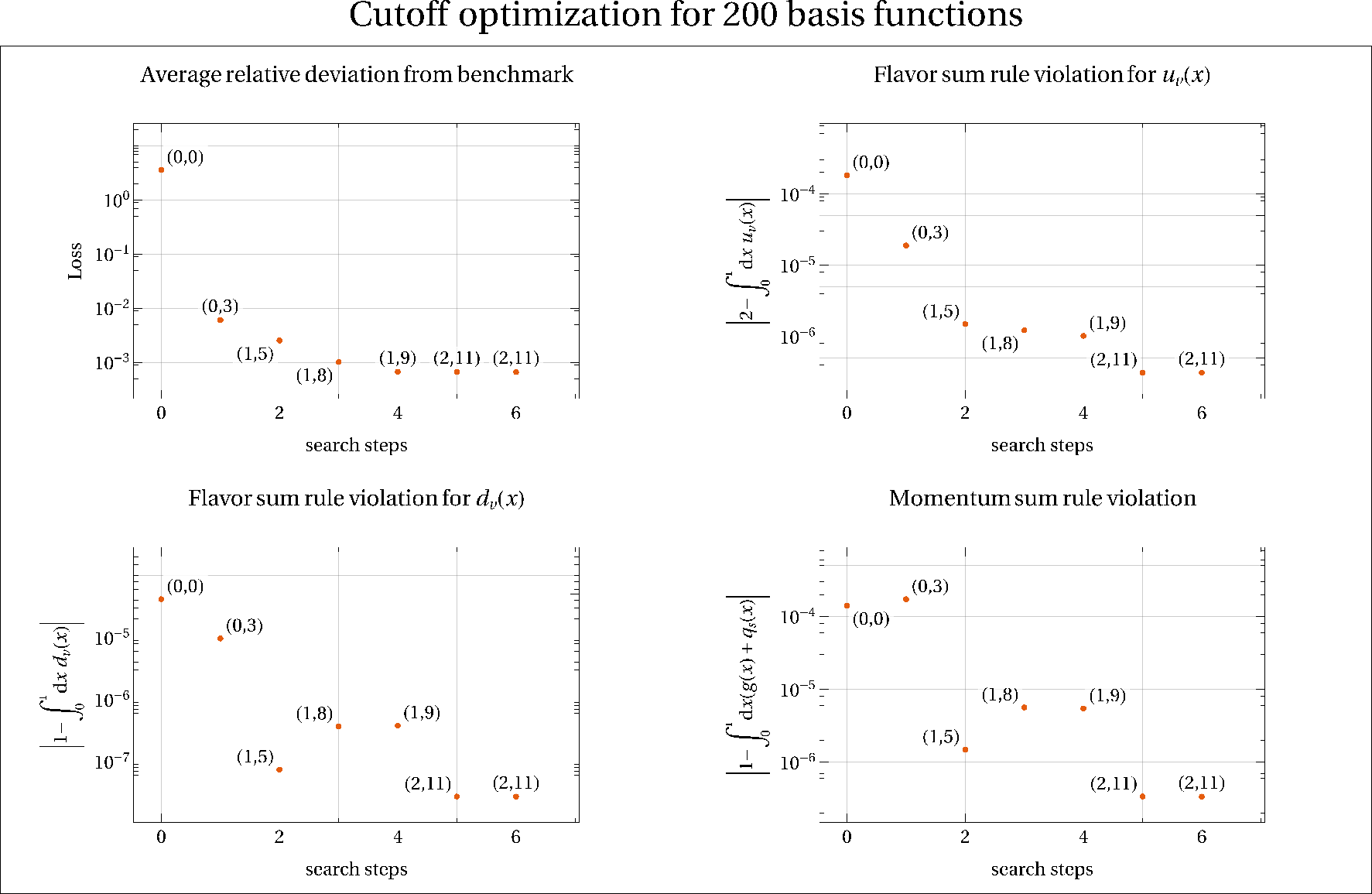}
		\caption{Plots comparing our method to the Les Houches benchmark tables for different cut-off functions \eqref{eq: Linear cutoff function}.
		The points are labelled with the corresponding cut-off parameter pairs $(\alpha,\beta)$. 
		The loss plotted in the upper left is the average deviation from the benchmark.
		Most of the difference stems from the very small sea-PDFs at $x=0.9$.
		The other three plots show the violation of sum rules as a metric for choosing a cut-off which does not rely on a pre-existing benchmark.}
		\label{fig: Cut off optimization Les Houches}
	\end{figure}
	
To determine the best cut-off, we optimize the parameters $\alpha$ and $\beta$ to minimize the deviation from the Les Houches benchmark tables \cite{Giele:2002hx}.
Starting from \change{the quadratic cut-off $(\alpha,\beta)=(0,0)$} we perform a gradient descent on the lattice of integer values for $\alpha$ and $\beta$ until we reach a local minimum.
This successive optimization of the cut-off is shown in figure \ref{fig: Cut off optimization Les Houches}.

Testing for $50$, $100$, $150$, and $200$ basis functions shows that the optimal cut-off parameters $\alpha$ and $\beta$ depend on the number of functions.
Starting from $100$ basis functions there is reasonable agreement for valence-like and sea-like distributions.
For $200$ basis functions most entries agree to all significant digits.
The exception to this are the very small sea-like distributions at $x=0.9$.
We optimized $\alpha$ and $\beta$ with respect to the average relative deviation of \texttt{POMPOM} and the benchmark tables.
This average is strongly dominated by these points.
Flavor and momentum sum rules are violated at a level of below $10^{-6}$.
This gives a quantitative measure for the global accuracy of our method.
We observe that optimizing the cut-off improves the $\texttt{POMPOM}$ method by some orders of magnitude.

The optimal cut-off may also depend on the initial conditions.
In the following, where we consider a different PDF set, for which no benchmark is available, we will use sum rule violations as a self-contained metric to optimize the cut-off.
As seen in figure \ref{fig: Cut off optimization Les Houches}, the cut-off which leads to the best agreement with the benchmark also has low sum rule violation.

\subsection{Evolution of realistic PDF set}
		\begin{figure}
		\centering
		\includegraphics[width=\textwidth]{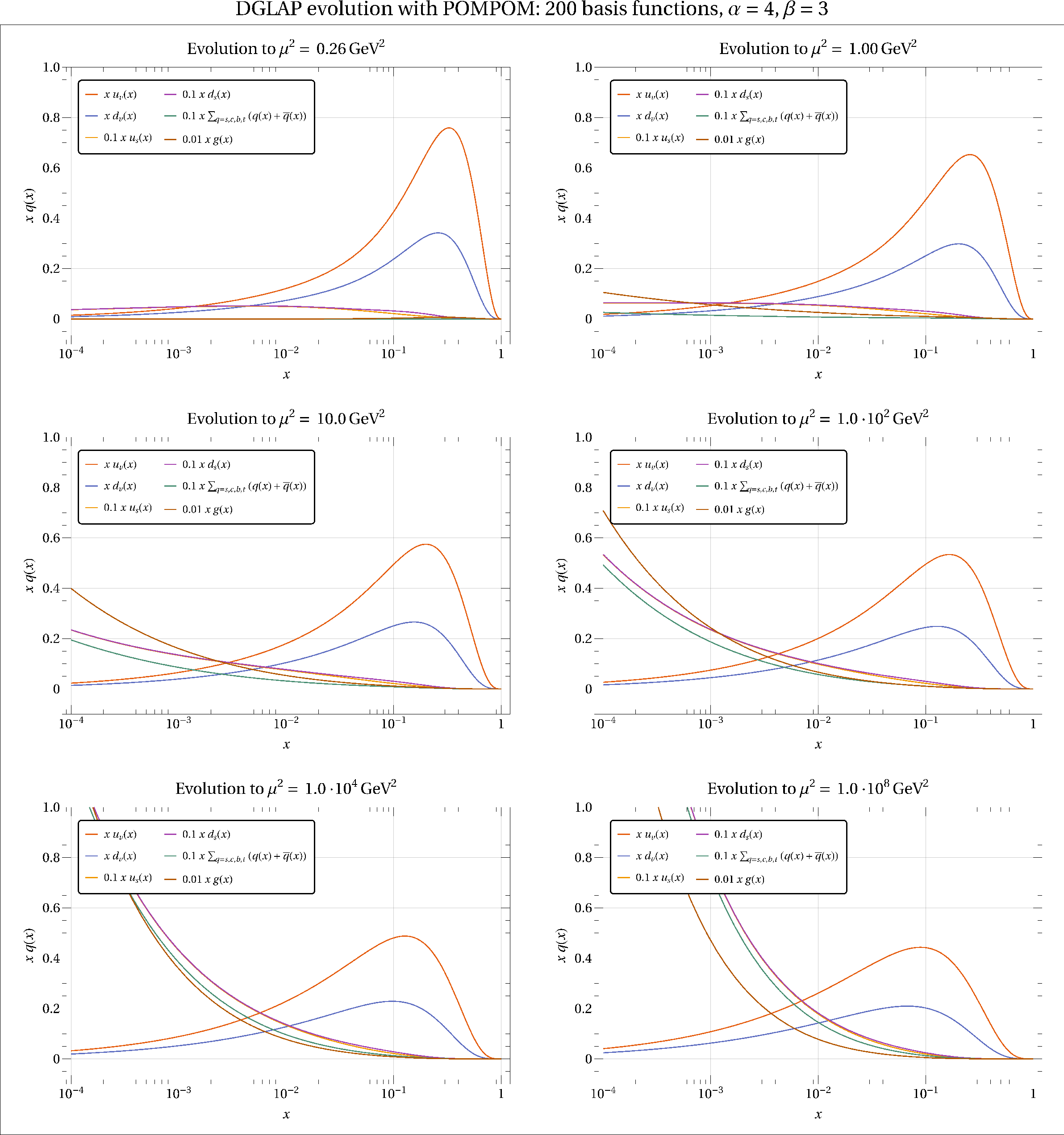}
		\caption{Evolution of a full set of PDFs in the variable flavor number scheme using \texttt{POMPOM}.
			For improved visibility, sea-like PDFs are scaled by a factor of $0.1$, the gluon PDF is scaled by $0.01$.}
		\label{fig: Full evolution}
	\end{figure}
From the comparison with the Les Houches benchmark tables we have seen that \texttt{POMPOM} works properly and preserves sum rules to a high level of accuracy.
To further showcase the capability of the proposed new method in the context of DGLAP evolution we change the initial conditions and turn to a more realistic model for PDFs.
For the numerical evolution in this section, we use the leading order input distributions at $\mu_0^2 = \unit[\change{0.26}]{GeV^2}$ from \cite{Gluck:1998xa}, which are listed in appendix \ref{app:Input_distributions} and take the running coupling at leading order, which is given by eq.\,\eqref{eq:alpha_s_LO}.
Note that while these input distributions have an analytic parametrization in $x$, discrete input distributions can be interpolated with functions of the form \eqref{eq:NonSinglet_DGLAP_ansatz_for_PDF_Cutoff} and subsequently evolved with the \texttt{POMPOM} method.
Technical details of such an interpolation are beyond the scope of the present work.
The evolution is performed in the variable flavor number scheme, where one evolves up to a flavor threshold, the resulting distribution is then taken as the new input above the threshold.
At leading order there are no non-trivial matching conditions\change{, which only come into play starting at NLO \cite{Bertone2014,Candido2022}.}
\change{Here, }the output distribution are immediately suitable to be new inputs without the need of additional steps.
This is in contrast to an evolution in Mellin-space where evolution space and output space do not coincide and a transformation is required.

The PDFs from \cite{Gluck:1998xa} provide us with sufficiently realistic initial conditions which are different from the benchmark to demonstrate that our method does not rely on specifics of the initial distributions.
We note that due to the change of initial conditions, the optimal cut-off is also slightly different.
As a way to fix the optimal cut-off in the absence of a previously known benchmark, we use the average \change{violation} of flavor and momentum sum rules.
As discussed in the comparison with the Les Houches benchmarks, these sum rules are a sensible measure for the global quality of the evolution.
	
The results for $200$ basis functions and cut-off parameters $\alpha=4$, $\beta=3$ are shown in figure \ref{fig: Full evolution}.
When the scale $\mu^2$ increases, the valence PDFs shift to smaller values of $x$ while preserving flavor number.
Simultaneously, the gluon and sea quark distributions strongly increase.
Hence, at larger scale, a larger fraction of the proton momentum is carried by the gluons and sea quarks, while the total momentum is preserved.

\subsection{Consistency of back-evolution}
\begin{figure}
		\centering
		\includegraphics[width=1\textwidth]{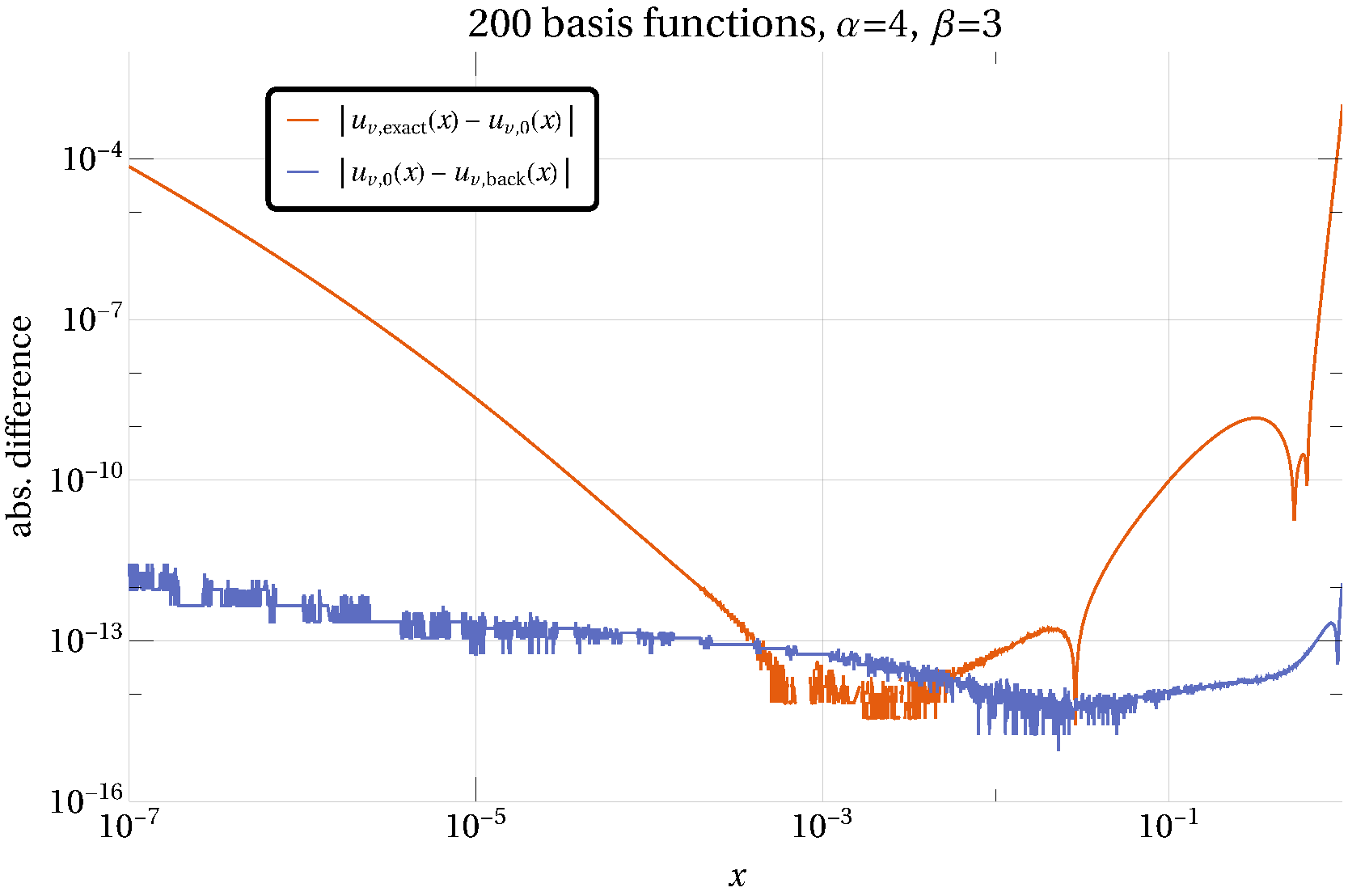}
		\caption{\change{Comparison of back-evolved PDF with initial condition.
		The red curve shows the difference between the exact initial condition $u_{v,\text{exact}}(x)$ and the initial condition expanded in a basis of 200 functions $u_{v,0}(x)$, which enters the evolution at $\mu_0^2=\unit[0.26]{GeV^2}$.
		The blue curve shows the difference between $u_{v,0}(x)$ and the back-evolved PDF $u_{v,\text{back}}(x)$, which has been evolved from $\mu_0^2=\unit[0.26]{GeV^2}$ to $\mu^2=\unit[10^8]{GeV^2}$ and back to $\mu_0^2=\unit[0.26]{GeV^2}$.}
		}
		\label{fig: Back Evolution}
	\end{figure}
To perform a consistency check of our evolution code, we evolve the initial PDFs from the starting scale $\mu_0^2=\unit[0.26]{GeV^2}$ to $\mu^2=\unit[10^8]{GeV^2}$ and then back to $\mu_0$.
Of course, for an exact evolution, the back-evolved PDF will exactly coincide with the initial PDF.
However, any approximate method may result in errors of the order of the approximations applied.
\change{In figure \ref{fig: Back Evolution} we show the difference between the initial condition expressed in the used basis $u_{v,0}(x)$ and the back-evolved PDF $u_{v,\text{back}}(x)$.
Compared to the difference due to truncation between the exact initial condition $u_{v,\text{exact}}(x)$ and the basis representation $u_{v,0}(x)$ we can consider the back-evolution to be exact up to numerical inaccuracies.
Absolute differences are plotted rather than relative ones to avoid division by very small values near $x=1$.}
  
In our case, due to the evolution employing an exponential, the evolution and back-evolution operators are exact inverses up to numerical inaccuracies.
Hence, truncation effects of the finite cut-off cancel between evolution and back-evolution.
Therefore, on the one hand, back-evolution is not suited to evaluate the numerical impact of cut-off effects, on the other hand exact back-evolution is a theoretically desirable feature.

\subsection{Comparison to Mellin-space evolution}
In this section, we compare \texttt{POMPOM} with evolution in Mellin-space.
The initial PDFs are again taken from \cite{Gluck:1998xa}.
For the Mellin evolution we use a custom \texttt{Mathematica} code based on the method presented in \cite{Vogt2005}.
The Mellin evolution code is included in the \texttt{Mathematica} version of \texttt{POMPOM} for comparison purpose.

\begin{figure}[t]
		\centering
		\includegraphics[width=\textwidth]{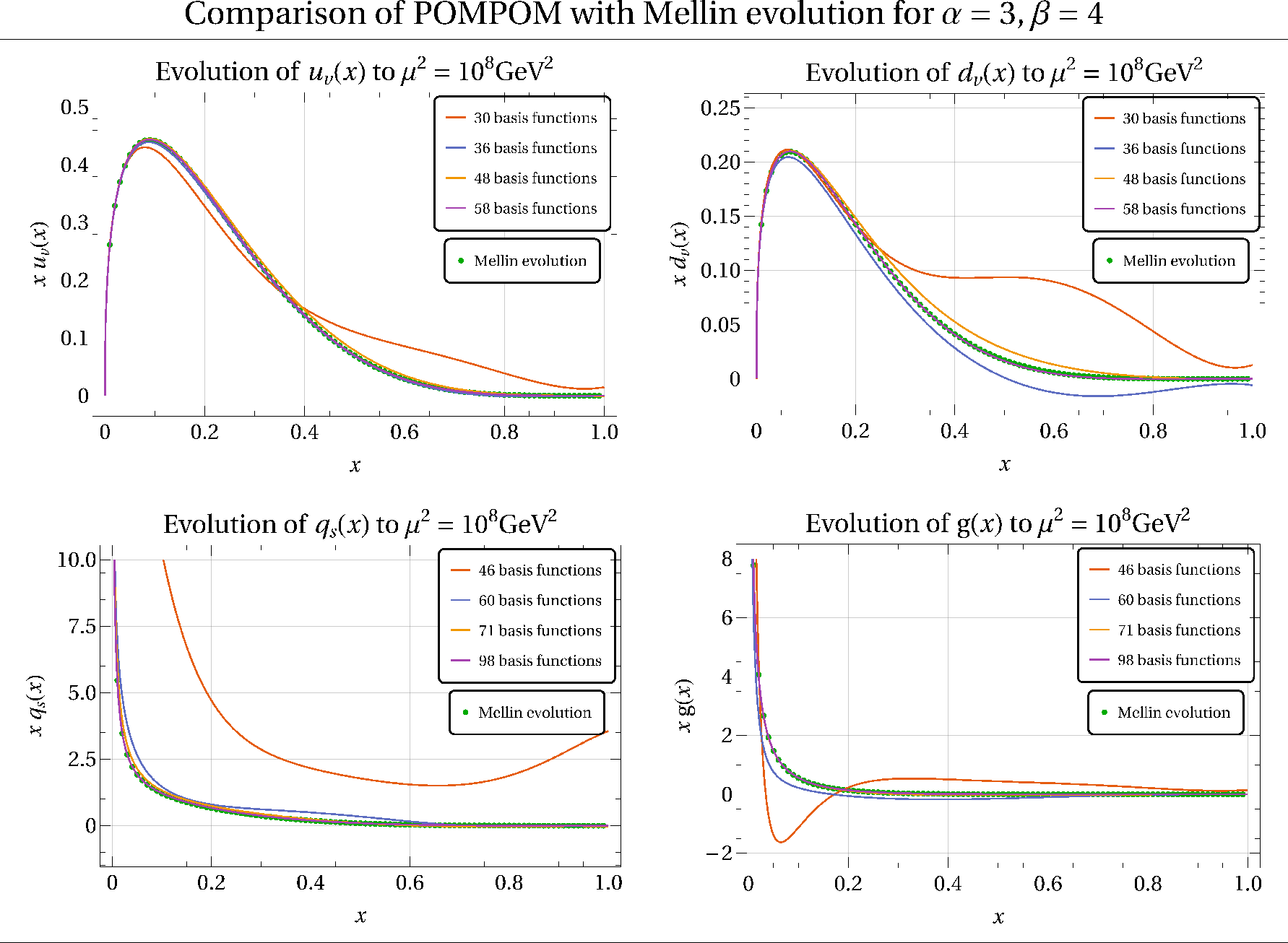}
		\caption{Comparison of \texttt{POMPOM} with Mellin-space evolution. For increasing numbers of basis functions, \texttt{POMPOM} results converge towards the Mellin results.}
		\label{fig: Mellin evolution}
\end{figure}
	
\change{First}, we consider rather small numbers of basis functions.
The idea is to investigate how many basis functions are required to obtain qualitatively plausible results.
This is potentially interesting if one aims for lowering computational cost on the expense of accuracy, which might be desirable when considering the evolution of multi-parton distributions in the future. There, the basis needs to span multiple dimensions, resulting in a increased number of required functions.
Results for the comparison of \texttt{POMPOM} with different numbers of basis functions to Mellin evolution are given in figure \ref{fig: Mellin evolution}.
	
The evolution up to $\mu^2=\unit[10^8]{GeV^2}$ is performed in the variable flavor number scheme.
The cut-off parameters are chosen as $\alpha=3$ and $\beta=4$, which are found to be the optimal numbers for $\sim 75$ basis functions.
\change{To showcase the qualitative features, absolute results are plotted on a linear scale.
A more detailed look at the level of agreement is taken below.}

We see that a sufficient number of basis functions is required to obtain qualitatively satisfactory results.
As long as basis functions essential for the description of the initial distribution are omitted from the basis, results are far off from the Mellin values.
This number is observed to be $\sim 50$ for valence-like distributions and $\sim 100$ for sea-like distributions.
This difference in required basis size is explained by the additional $\frac{1}{x}$ terms to describe stronger divergence of the sea distributions at small-$x$. 

\begin{figure}[t]
		\centering
		\includegraphics[width=1\textwidth]{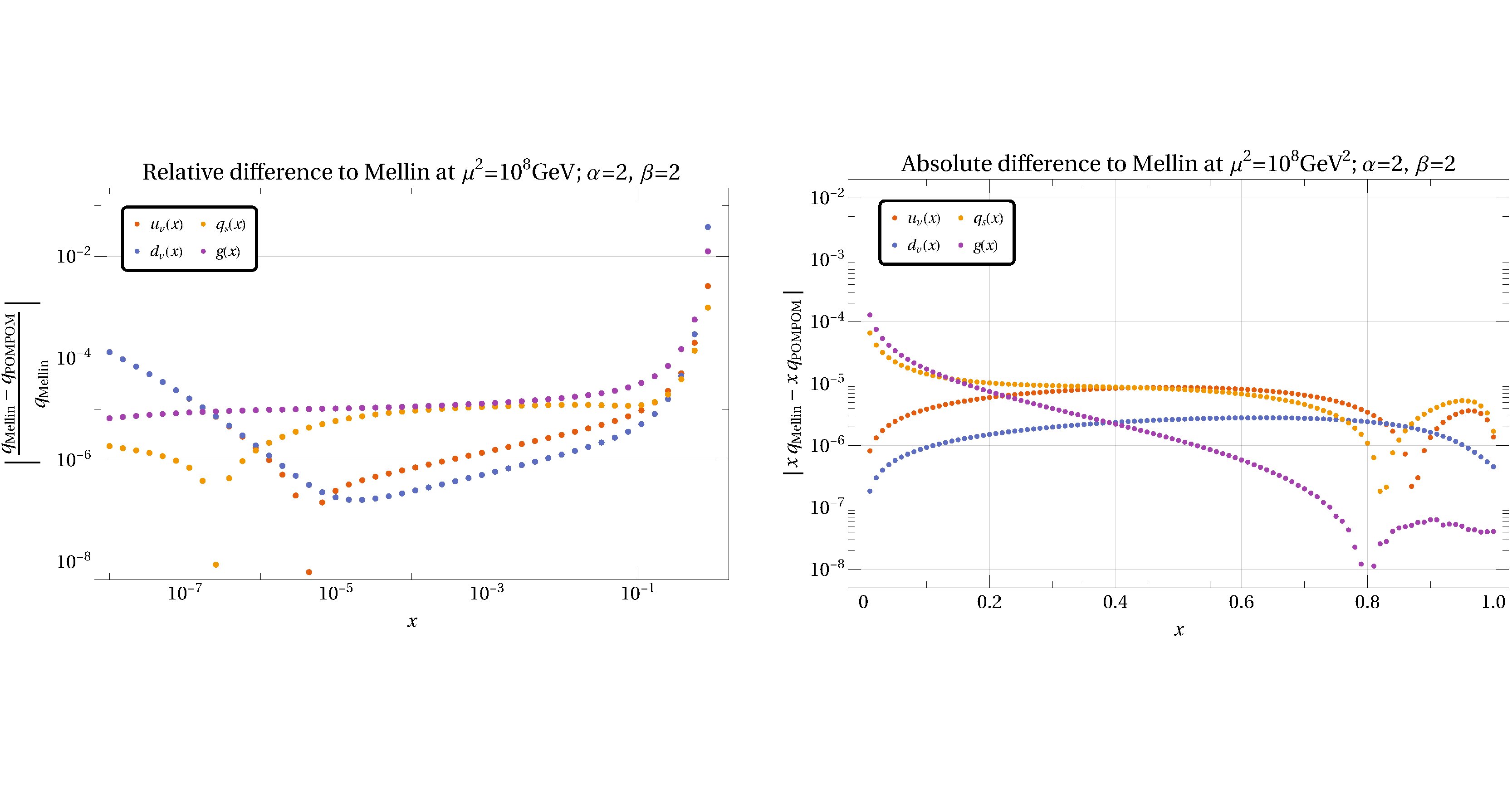}
		\vspace{-1cm}
		\caption{\change{Relative and absolute differences between \texttt{POMPOM} and Mellin \change{space} evolution for non-singlet and singlet evolution.
		The left figure shows the relative differences on a logarithmic $x$-scale to visualize small-$x$ behavior, the right figure shows absolute differences on a linear $x$-scale to visualize large $x$ behavior.}}
		\label{fig: Mellin Evolution 2}
	\end{figure}
	
\change{Now we take a closer look at the level of agreement between \texttt{POMPOM} and Mellin evolution.
For this purpose we take a moderately higher number of $200$ basis functions with cut-off parameters $(\alpha,\beta)=(2,2)$.
Evolution is performed to $\mu^2=\unit[10^8]{GeV^2}$ as above, the relative and absolute differences between \texttt{POMPOM} and Mellin evolution are plotted in figure \ref{fig: Mellin Evolution 2}.
To visualize the small-$x$ behavior, the left figure shows relative differences on a logarithmic scale.
For the non-singlet evolution we find agreement below the $\sim 10^{-5}$ level for $10^{-7}\lesssim x\lesssim 0.2$ with minimal relative difference around $x=10^{-4}$.
For the singlet evolution the agreement is on this level for $x\lesssim 0.1$ and is roughly constant down to $x=10^{-8}$.
In the right figure, we look at the large-$x$ behavior on a linear $x$-scale.
To avoid division by the small values of PDFs near $x=1$ absolute differences between \texttt{POMPOM} and Mellin Evolution are presented in this case.
For large $x$ the absolute differences stay below $10^{-5}$ and do not increase when approaching $x=1$ confirming the findings from the comparison with the Les Houches benchmark in figure \ref{fig: Comparison with Les Houches}.
}

\change{These comparisons show that \texttt{POMPOM} is in reasonable agreement with well-established methods for DGLAP evolution at leading order, demonstrating that the cut-off effect can be controlled with a numerically manageable number of basis functions.
Since higher orders in $\alpha_s$ will presumably, due to their suppression in $\alpha_s$, only result in moderate corrections to the matrix elements used in \texttt{POMPOM} evolution, one can hope for a comparable level of numerical agreement for a certain number of basis functions.
A detailed comparison to existing evolution codes such as \texttt{HOPPET}, \texttt{APFEL}, and \texttt{EKO} is beyond the scope of this paper since it will be meaningful only after \texttt{POMPOM} has been upgraded to a state-of-the-art DGLAP evolution code.
Here we only aim for a proof-of-concept of the proposed method to justify its future developments for higher order DGLAP evolution or structurally similar evolution equations.
}

\subsection{Convergence for increasing number of basis functions}
In this section, we take a closer look at the convergence of \texttt{POMPOM} when increasing the number of basis functions.
It is clear that increasing the number of basis functions improves the result.
However, how quick the convergence towards the true value turns out to be requires investigation.

When comparing with Mellin-space evolution in the previous section, we have already seen some quantitative examples of convergence.
Here, we take a closer look at the quark singlet distribution, which is especially difficult to describe due to its strong divergence for small $x$ and particularly its very small values for $x\approx 1$.

We fix the cut-off parameters at $\alpha=\beta=2$ and perform an evolution to $\mu^2=\unit[10^8]{GeV^2}$ with the PDF set from from \cite{Gluck:1998xa} while successively increasing the number of basis functions.
Then we compare the evolved PDFs for different numbers of basis functions.

	\begin{figure}[t]
		\centering
		\includegraphics[width=0.87\textwidth]{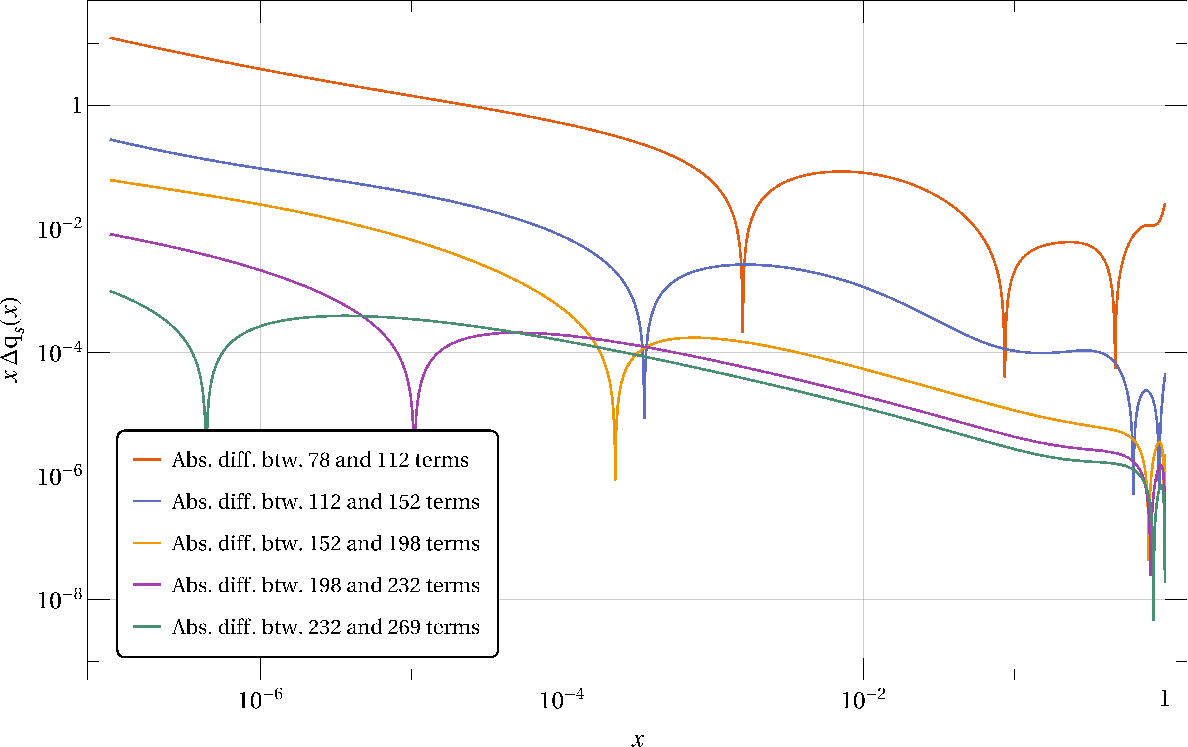}
		\caption{Absolute difference between singlet PDFs $x\, q_{\text{s}}(x)$ evolved from scale $\mu_0^2 = \unit[\change{0.26}]{GeV^2}$ to $\mu^2 = \unit[10^8]{GeV^2}$ with \texttt{POMPOM} for increasing number of basis functions. The cut-off parameters are $\alpha=\beta=2$.}
		\label{fig: Convergence plot absolute differences}
	\end{figure}
	
\begin{figure}[t]
		\centering
		\includegraphics[width=0.87\textwidth]{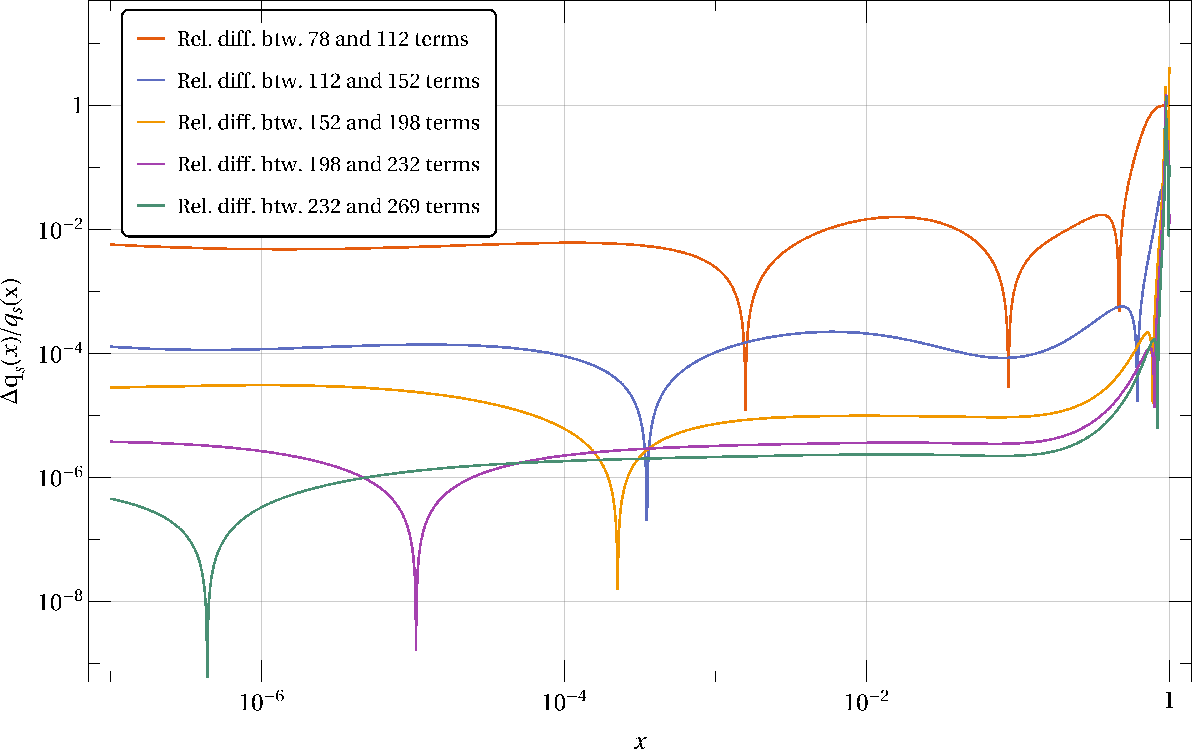}
		\caption{Relative difference between singlet PDFs $x\, q_{\text{s}}(x)$ evolved from scale $\mu_0^2 = \unit[\change{0.26}]{GeV^2}$ to $\mu^2 = \unit[10^8]{GeV^2}$ with \texttt{POMPOM} for increasing number of basis functions. The cut-off parameters are $\alpha=\beta=2$.}
		\label{fig: Convergence plot relative differences}
	\end{figure}

In figures \ref{fig: Convergence plot absolute differences} and \ref{fig: Convergence plot relative differences}, respectively, we compare the absolute and relative differences for evolutions with between $78$ and $269$ basis functions in steps of $\sim 40$.
We observe that the differences between steps becomes smaller and smaller indicating convergence towards zero.
The absolute differences are largest for small $x$ since the singlet PDF grows strongly in this region.
The lowest absolute differences occur at large $x\approx 1$.
The relative differences are roughly constant for many orders of magnitude of $x$ between $10^{-7}$ and $10^{-1}$.
For large $x\approx 1$ the relative difference deteriorates due to the smallness of the singlet PDF in this region.
Overall, we observe satisfactory convergence for a modest linear increase in the number of basis functions.
By adjusting the cut-off for higher numbers of basis functions, the rate of convergence could be further improved.
\section{Conclusion and Outlook}\label{sec:Conclusion}
We have demonstrated that the proposed $x$-space method implemented in \texttt{POMPOM} is capable of accurately solving the DGLAP equation for the non-singlet and singlet cases.
We obtained good agreement with both the Les Houches benchmark set as well as \change{with Mellin-space evolution of} the PDFs from \cite{Gluck:1998xa}.
Sum rule violations drop to levels below $10^{-6}$ for $\mathcal{O}(200)$ basis functions, relative corrections when further increasing the number of basis functions are below $10^{-4}$ except very close to $x=1$.
\texttt{POMPOM} is accurate over a wide range of $x$ as well as $\mu$, we tested for $2\,\Lambda_\text{QCD}\simeq\mu<\unit[10]{TeV}$ and $10^{-7}<x<1$.
The largest deviations are observed for $x\approx 1$ and originate from truncating of the series for $\ln(1-x)$.
One could try to improve accuracy in the large-$x$ region by amending the basis by $\ln(1-x)$, however since we achieved a, for our purposes, satisfactory level of accuracy, this is beyond the scope of the present paper.
\change{The performed numerical studies demonstrate that for the leading order DGLAP equation the cut-off effect of the \texttt{POMPOM} method can indeed be controlled with a numerically manageable number of basis functions.
This is an essential foundation for a potential development of \texttt{POMPOM} towards an state-of-the-art DGLAP evolution program that could for example be used for PDF fitting.
}

Structurally, the DGLAP equations are very similar to the ETQS evolution equation of the twist-3 Qui-Sterman function.
In the singlet case the latter is \cite{Pirnay:2013}
	\begin{align}
		\frac{\partial}{\partial\ln\mu^2}\left(\begin{array}{c}
		\mathfrak{S}^\pm
		\\
		\mathfrak{F}^\pm
		\end{array}\right)=
		-\frac{\as(\mu^2)}{2\pi}\left(\begin{array}{cc}
		\mathbb{H}^\pm_{QQ}&\mathbb{H}^\pm_{QF}
		\vspace{2pt}
		\\
		\mathbb{H}^\pm_{FQ}&\mathbb{H}^\pm_{FF}
		\end{array}\right)
		\left(\begin{array}{c}
		\mathfrak{S}^\pm
		\\
		\mathfrak{F}^\pm
		\end{array}\right) ,
		\label{eq: TF evolution}
	\end{align}
where $\mathfrak{S}^\pm$ and $\mathfrak{F}^\pm$ are functions of two momentum fractions.
The $\mathbb{H}^\pm$ are linear one-dimensional integral operators.
Expressions for them can be found in the appendix of \cite{Braun2009}. 
The non-singlet case decouples similar to DGLAP evolution.
The existing evolution code \texttt{t3evol} \cite{Pirnay:2013} for eq.\,\eqref{eq: TF evolution} uses discretization in both $\mu$ and $x$-space.\footnote{\change{After the publication of the first preprint of the present paper the new twist-3 evolution codes \texttt{Honeycomb}/\texttt{Snowflake} appeared \cite{Rodini:2024}. These use an improved discretization approach.}}
This code was  for example applied in \cite{Hatta:2019}.

Having established the validity of the $x$-space method proposed in this paper for the DGLAP case, generalizing \texttt{POMPOM} to the evolution equation \eqref{eq: TF evolution} is a logical next step.
Here, finding a suitable set of basis functions spanning the two-dimensional space of momentum fractions will be key.
Along the lines of the presented work, one may start from a polynomial ansatz in both momentum fraction variables and determine which classes of functions are generated by the integration kernels.
These functions are subsequently incorporated in the basis if they are not well represented by truncated power series.

From there, selecting an optimal finite subset of basis functions can be expected to be somewhat more challenging than the DGLAP case.
Presumably, since there are two momentum fraction variables, more functions will be required for accurate description of the distributions.
Hence, it will be more important to find optimal small sets of basis functions to lower the required number. 
As we have seen in the DGLAP case, for a fixed number of basis functions, the choice of cut-off can lead to order-of-magnitude improvements in accuracy.

An idea worth investigating might be the use of machine learning techniques for the task of cut-off optimization.
As a starting point, the impact of a cut-off fitted by a neural net for the DGLAP case could be assessed by comparing with the two-parameter cut-off used in this work.

\acknowledgments
We thank Dominik Bammert for inspiring this work.
We are grateful to Werner Vogelsang for useful comments during all stages of this work.
\change{We also thank Andrea Simonelli for useful discussions regarding the application of the Magnus expansion.}
The authors acknowledge helpful comments by Johannes Bl\"umlein regarding the literature of state-of-the-art calculations of QCD splitting functions and the running coupling.
This work has been supported by Deutsche Forschungsgemeinschaft (DFG) through the Research Unit FOR 2926 (project 409651613).
J.H. is grateful to the Landesgraduiertenf\"orderung Baden-W\"urttemberg for supporting her research.

%
%
\appendix 

\section{Master integrals}\label{app:Master_Integrals}
In this appendix, we list the results for the master integrals appearing in eq.\,\eqref{eq:NonSinglet_DGLAP_I1_I2}.
First, we have
	\begin{align}
		I_1^{m,n} \,&\equiv\, x^n \int_x^1 \dx\xi\, \frac{\left(\xi^{-n} - 1\right)}{1-\xi} \frac{\ln^m(x/\xi)}{m!} &
		\\
		&=\, \left\{
		\begin{array}{ll}
			-\sum_{k=0}^{-n-1} \left[ \sum_{j=0}^m \frac{1}{(k+1)^{m+1-j}}\; \frac{\ln^j(x)\,x^n}{j!}  \,-\, \frac{x^{k+n+1}}{(k+1)^{m+1}} \right] &\; \text{if } n \,\leq\, -2 \,,
			\\
			-\frac{1}{x}\sum_{k=0}^m \frac{\ln^k(x)}{k!} \,+\, 1 &\; \text{if } n \,=\, -1 \,,
			\\
			0 &\; \text{if } n \,=\, 0 \,,
			\\
			x^n \left[ (-1)^m \sum_{k=1}^{n-1} \frac{x^{-k}}{k^{m+1}} \,-\, \sum_{k=0}^m (-1)^{k}\, H_{n-1,k+1}\, \frac{\ln^{m-k}(x)}{(m-k)!} \,-\, \frac{\ln^{m+1}(x)}{(m+1)!}\right]  &\; \text{if } n \,\geq\, 1 \,.
		\end{array}
		\right. 
	\end{align}
The second master integral is given by
	\begin{align}
		I_2^{m,n} \,&\equiv\, x^n \int_x^1 \frac{\dx\xi}{(1-\xi)}\, \frac{\ln^m(x/\xi) \,-\, \ln^m(x)}{m!}
		\\
		\,&=\, x^n \left[ \sum_{k=0}^{m-1} \zeta_{m-k+1} \frac{\ln^k(x)}{k!} \,+\, \frac{\ln^m(x)}{m!}\sum_{k=1}^{\infty} \frac{x^k}{k} \,-\, \sum_{k=1 }^{\infty} \frac{x^k}{k^{m+1}} \right] .
	\end{align}
For the singlet DGLAP equation, we also need the third master integral
	\begin{align}
		I_ 3^{m,n} \,&\equiv\,  x^n \int_x^1 \dx\xi\, \frac{\ln^m(x/\xi)\, \xi^{-n}}{m!}
		\\
		&=\, \left\{
		\begin{array}{ll}
			\sum_{j=0}^m \frac{1}{(1-n)^{m+1-j}} \frac{\ln^j(x)\, x^n}{j!} \,-\, \frac{x}{(1-n)^{m+1}} &\quad \text{if } n \,\neq\, 1 \,,
			\\
			- \frac{\ln^{m+1}(x)\, x}{(m+1)!} &\quad \text{if } n \,=\, 1 \,.
		\end{array} 
		\right.
	\end{align}
Here, $H_{n,m} \equiv \sum_{k=1}^{n}\frac{1}{k^m}$ are the generalized harmonic numbers, while $\zeta_m \equiv \sum_{k=1}^{\infty}\frac{1}{k^m}$ are the integer values of the Riemann zeta function.

\section{Differential equation for LO non-singlet coefficients}\label{app: Differential equation for LO non-singlet coefficients}
Explicitly inserting the master integrals given in appendix \ref{app:Master_Integrals} into eq.\,\eqref{eq:NonSinglet_DGLAP_I1_I2} and collecting the result with respect to the spanning functions yields the following differential equation for the coefficients
	\begin{align}
		\frac{\dx\, a_{MN}\!\left(\mu^2\right)}{\dx \ln\mu^2} \,=\, \frac{\as\!\left(\mu^2\right) \CF}{2\pi} &\left[\,\sum_{m>M}\, \alpha(N,m-M) a_{mN} \,+\, 	\beta(N)\, a_{MN} \right.
		\nonumber \\
		&\left. +\, \gamma(N)\left(\delta_{M0} \sum_{\substack{m,n\\n\neq N}} \frac{a_{mn}}{(N-n)^{m+1}} \,+\, a_{M-1,N}\right) \right],
		\label{eq:NonSinglet_DGLAP_for_coefficients_LO_explicit}
	\end{align}
where
	\begin{align}
		\alpha(N,l) \,\equiv \left\{
		\begin{array}{ll}
			2\zeta_{l+1} \,-\, 1 &\quad \text{if } N \,=\, 0 \,,
			\\
			2\zeta_{l+1} \,-\, (-1)^{l} &\quad \text{if } N \,=\, 1 \,,
			\\
			2\zeta_{l+1} \,-\, (-1)^{l} \left( H_{N,l+1} \,+\, H_{N-2,l+1} \right) &\quad \text{if 	} N \,\geq\, 2 \,,
		\end{array}
		\right. 
	\end{align}
and
	\begin{align}
		\beta(N) \,\equiv \left\{
		\begin{array}{ll}
			\frac{1}{2} &\quad \text{if } N \,=\, 0, 1 \,,
			\vspace{2pt}
			\\
			\frac{3}{2} \,-\, H_N \,-\, H_{N-2} &\quad \text{if } N \,\geq\, 2 \,,
		\end{array}
		\right.
	\end{align}
and 
	\begin{align}
		\gamma(N) \,\equiv \left\{
		\begin{array}{ll}
			-1 &\quad \text{if } N \,=\, 0, 1 \,,
			\\
			-2 &\quad \text{if } N \,\geq\, 2 \,.
		\end{array}
		\right.
	\end{align}
Here, $H_{n,m} \equiv \sum_{k=1}^{n}\frac{1}{k^m}$ are the generalized harmonic numbers, and $H_n \equiv H_{n,1} = \sum_{k=1}^{n}\frac{1}{k}$ the usual harmonic numbers.
The matrix $\mathcal{P}$ introduced in eq.\,\eqref{eq:NonSinglet_DGLAP_for_coefficients} can be read off from eq.\,\eqref{eq:NonSinglet_DGLAP_for_coefficients_LO_explicit}.

\section{Input distributions and parameters}\label{app:Input_distributions}
For numerical comparison of our DGLAP evolution with Mellin-space evolution, we use the sufficiently realistic leading order input distributions from \cite{Gluck:1998xa} at $\mu_0^2 = \unit[0.26]{GeV^2}$,
	\begin{align}
		x\, u_v\!\left(x,\mu_0^2\right) &=\, 1.239\, x^{0.48}\, (1-x)^{2.72} \left(1 - 1.8\sqrt{x} + 9.5 x\right) ,
		\\
		x\, d_v\!\left(x,\mu_0^2\right) &=\, 0.614\, (1-x)^{0.9}\, x\, u_v\!\left(x,\mu_0^2\right) \,,
		\\
		x\,\Delta\!\left(x,\mu_0^2\right) &\,=\, 0.23\, x^{0.48}\, (1-x)^{11.3} \left(1-12.0\sqrt{x} + 50.9 x\right) ,
		\\
		x\, (\bar{u}+\bar{d})\!\left(x,\mu_0^2\right) &=\, 1.52\, x^{0.15}\, (1-x)^{9.1} \left(1 - 3.6 \sqrt{x} + 7.8 x \right) ,
		\\
		x\, g\!\left(x,\mu_0^2\right) &=\, 17.47\, x^{1.6}\, (1-x)^{3.8} \,,
		\\
		x\, s\!\left(x,\mu_0^2\right) &=\, x\, \bar{s}\!\left(x,\mu_0^2\right) \,=\, 0 \,,
	\end{align}
where $\Delta \equiv \bar{d} - \bar{u}$. At leading order, the running coupling is given by
	\begin{align}
		\as\!\left(\mu^2\right) =\, \frac{4\pi}{\beta_0 \change{\,\ln\!\left(\frac{\mu^2}{\Lambda_{\text{QCD}}^2}\right)}}\,,
		\label{eq:alpha_s_LO}
	\end{align}
where $\beta_0 \equiv 11 - \frac{2}{3}\Nf$, while the QCD mass scale is ${\Lambda_{\text{QCD}}^{(\Nf = 3,4,5,6)} = 204, 175, 132, \unit[66.5]{MeV}}$.
The flavor number thresholds for $\Nf=4,5,6$ are at $\mu^2=1.96, 20.25, \unit[30625]{GeV^2}$.
\section{Computation time}
\label{app: Computation time}
\begin{figure}[t]
\vspace{-0.8cm}
		\centering
		\includegraphics[width=1\textwidth]{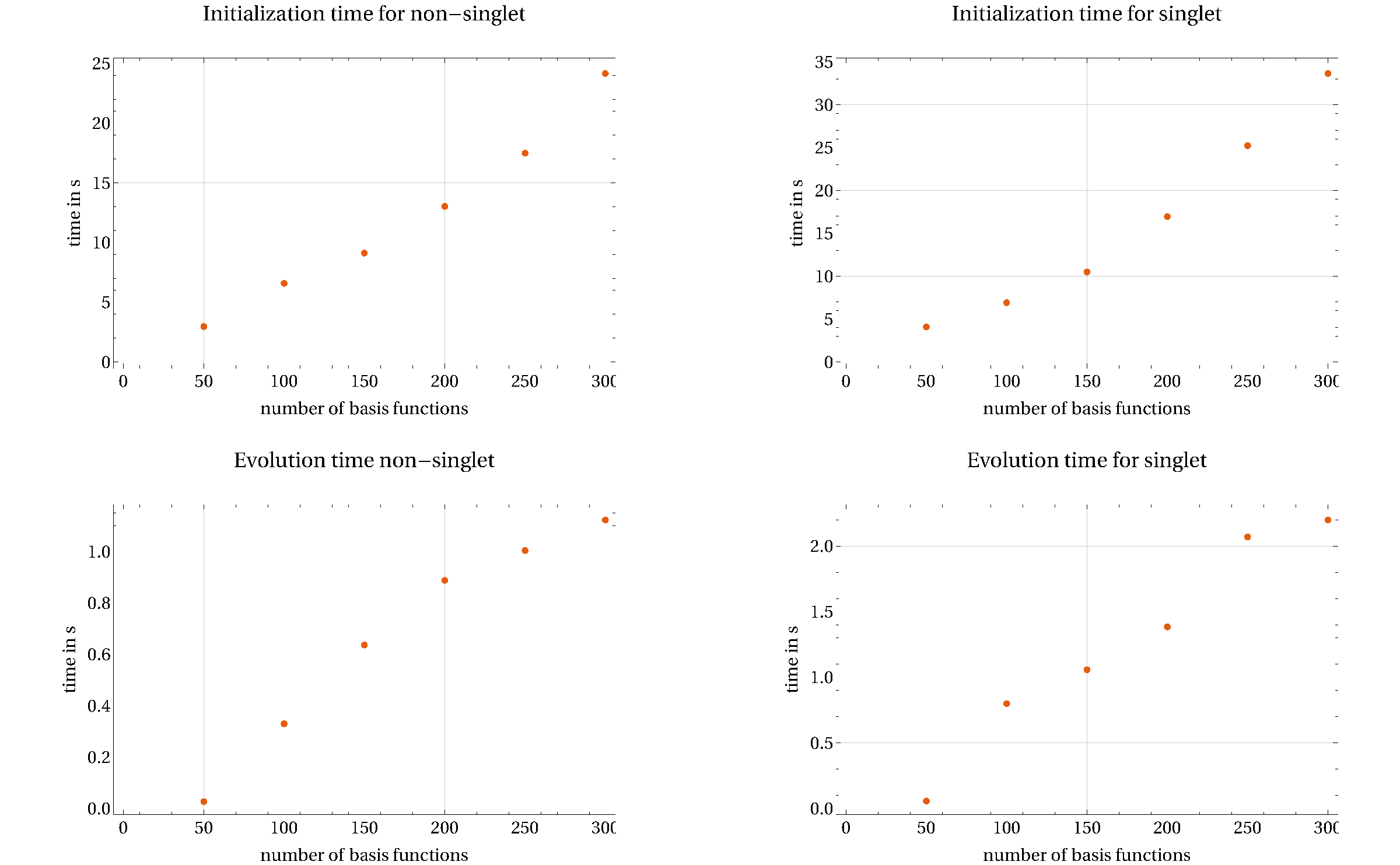}
		\caption{\change{Time required for initialization and evolution of non-singlet and singlet evolution. The timing was performed with \texttt{Mathematica 13.1} on a ThinkPad T495 with $\unit[8]{GB}$ of RAM.}}
		\label{fig: Timing}
	\end{figure}
\change{Since the current version of \texttt{POMPOM} is mainly designed for explorative purposes, time performance is not considered the most important aspect of the code.
For a potential future upgrade of \texttt{POMPOM} into a code applicable to PDF fits this will be of relevance.
One also has to keep in mind that a comparison between evolution programs is only meaningful in reference to a specified task as was dicussed in \cite{Candido2022}, where the timing of \texttt{EKO} is compared with \texttt{APFEL}.
}

\change{
To get a first estimate on the time required for PDF evolution with \texttt{POMPOM}, we show initialization times and the time for the evolution to a single scale for non-singlet and singlet evolution in figure \ref{fig: Timing}.
Initialization has to be performed only once for a fixed number of basis functions with a certain cut-off to calculate the corresponding evolution matrix.
The initialization time grows approximately quadratically with the size of the basis.
Due to reuse of already calculated matrix elements for lower number of basis functions, the initialization with higher number of basis functions shown in the upper two plots is slightly quicker than it would be in a standalone calculation.
The subsequent evolution to a single scale is much faster in comparison and only grows linearly with increasing number of basis functions.
For a reasonable number of basis functions, the evolution to a specific scale takes on the order of $\unit[1]{s}$ on a standard consumer laptop.
}

%
%
\bibliography{Bibliography_semi_analytical_solution_for_parton_evolution}
\bibliographystyle{JHEP}

\end{document}